\documentclass[a4paper,11pt]{article}
\pdfoutput=1
\usepackage{jcappub}
\usepackage[T1]{fontenc}
\usepackage{graphicx}
\usepackage{todonotes}
\usepackage{changepage}
\usepackage{academicons}
\usepackage{natbib}
\usepackage{multirow}
\usepackage{orcidlink}
\usepackage{subcaption}
\usepackage{soul}
\usepackage{xcolor}
\usepackage{amsmath}
\usepackage{booktabs}
\usepackage{siunitx}
\usepackage{lineno}
\usepackage{tabularx}
\usepackage{array}
\usepackage{threeparttable}
\usepackage{float}
\usepackage{longtable}
\newcolumntype{Y}{>{\centering\arraybackslash}X}
\definecolor{orcidlogocol}{HTML}{A6CE39}
\usepackage{aas_macros}

\title{\boldmath Comparative Periodogram Analysis of 22 Years of Super-Kamiokande Solar $^{8}\mathrm{B}$ Neutrino Data: Classical, Phase-Based, and Information Theoretic Methods}

\author[a]{Liangliang Ren\orcidlink{0000-0002-1428-4003},
\thanks{Corresponding author.}}
\author[a]{Ze-Lin Zhang\orcidlink{0000-0003-4621-0807}
\thanks{Corresponding author.}}
\author[a]{Bing Xu\orcidlink{0000-0002-9394-0426},
\thanks{Corresponding author.}}
\author[a]{Tian-Cheng Huang\orcidlink{0009-0001-2036-5141}}
\author[a]{Ran Wang\orcidlink{0009-0006-5768-1793}}
\author[a]{Jia-Xin Dong\orcidlink{0009-0004-0801-9993}}
\author[a]{Yan-Ping Wang\orcidlink{0009-0003-6805-2974}}

\affiliation[a]{School of Electrical and Electronic Engineering, Anhui Science and Technology University, Bengbu, Anhui 233030, People's Republic of China}

\emailAdd{rll@ahstu.edu.cn}
\emailAdd{zhangzl@ahstu.edu.cn}
\emailAdd{xub@ahstu.edu.cn}

\abstract{
Solar $^8\mathrm{B}$ neutrinos offer a unique probe of solar interior dynamics and neutrino electromagnetic properties. We present a systematic, multi-method periodogram analysis of the 22-year Super-Kamiokande solar neutrino dataset (1996--2018), comparing nine algorithms spanning classical, generalized, phase-dispersion, and information-theoretic paradigms. Through hierarchical temporal segmentation, we disentangle astrophysical signals from detector-specific systematics. We demonstrate that the Generalized Lomb-Scargle (GLS) method provides the most statistically robust detections by correctly handling heteroscedastic uncertainties, whereas the classical Lomb-Scargle method systematically underestimates significance. The Lafler--Kinman String Length (LKSL) method generally fails to yield effective detections, and when it does, its best-fit period deviates significantly from the consensus. Crucially, independent algorithms like Multi-Harmonic Analysis of Variance (MHAOV) and Phase Dispersion Minimization (PDM1) recover consistent periodicities, providing vital cross-validation.
In the pre-2001 and SK-I data, seven algorithms provide \textit{weak evidence} ($\ln B > 0$) for a $\sim 0.106$ yr ($\sim 38.8$ d) periodicity. However, this signal is entirely absent in the highest-statistics SK-IV modified flux data, where the Bayes factor decisively favors the null model ($\ln B \ll -5$), indicating it is a transient feature of the early low-statistics era. Conversely, a $\sim 24.3$ d signal in the post-2001 raw flux is decisively rejected by the Bayesian framework and vanishes in the modified flux, confirming its seasonal systematic origin. Furthermore, no evidence is found for an $\sim 11$-year solar cycle modulation, yielding a stringent amplitude upper limit of $<0.2\%$ of the mean flux.
By highlighting the stark contrast between frequentist significance (FAP) and Bayesian model selection ($\ln B$) in low signal-to-noise regimes, we establish a rigorous, multi-metric best-practice framework for periodicity searches. This work provides a direct methodological blueprint for next-generation observatories like Hyper-Kamiokande and JUNO.
}

\arxivnumber{2607.27979}

\begin{document}

\maketitle
\flushbottom

\section{Introduction}
\label{sec:introduction}

Solar $^8\mathrm{B}$ neutrinos, produced in the high-temperature core of the Sun via the $pp$-chain, serve as a direct probe of solar fusion dynamics and neutrino quantum properties~\cite{SNO:2001kpb,Adelberger:2010qa,Suliga:2021JC}. Their production rate scales steeply with the core temperature ($\Phi \propto T_c^{25}$), making them exceptionally sensitive to internal solar perturbations~\cite{Bahcall:1995bt,Bahcall:1998jt,Bahcall:2004pz}. As these neutrinos propagate outward, their flavor composition is modified by the Mikheyev--Smirnov--Wolfenstein (MSW) matter effect, which has been precisely validated by decades of real-time detection in water Cherenkov, liquid scintillator, and liquid xenon experiments~\cite{Super-Kamiokande:1998qwk,Borexino:2008fkj,XENON:2024ijk}.

Beyond standard oscillations, the solar $^8\mathrm{B}$ flux may carry additional temporal imprints if neutrinos possess non-standard electromagnetic properties or if the solar interior exhibits dynamic magnetic/rotational structures~\cite{Akhmedov:1988uk,Lim:1987tk}. Specifically, if neutrinos carry a non-zero magnetic moment (Dirac or Majorana type), resonant spin-flavor precession (RSFP) in the solar tachocline or radiative zone could periodically convert active $\nu_e$ into sterile or right-handed states, thereby modulating the detected elastic-scattering or coherent-scattering rates~\cite{Lim:1990Ph,Ando:2003Ph,Joshi:2020RA,Delepine:2026uph,Delepine:2026ar}. Furthermore, differential rotation profiles, magnetic dynamo cycles, quasi-biennial oscillations (QBO), subtle core temperature fluctuations, or even the collective effects of solar gravity (g) modes could imprint characteristic periodicities on the emergent neutrino flux~\cite{Aharmim:2010Ap,Baldwin:2001,Simoniello:2012pn,Sturrock:2022dgy,Vasil:2024,Hatta:2026gmode}.

Time-series periodicity analysis thus sits at a critical intersection of solar astrophysics and particle physics. Detecting or constraining sub-percent modulations in the $^8\mathrm{B}$ flux provides a unique window into the otherwise inaccessible solar radiative interior, complements helioseismic constraints on core rotation and solar gravity (g) modes, and places stringent bounds on neutrino magnetic moments and non-standard interactions (NSI)~\cite{Turck:2011RP,Zavatarelli:2020,Hatta:2026gmode}. The statistical power of such searches has dramatically increased with multi-decade datasets: Super-Kamiokande (SK) has accumulated over 5,800 live days across four operational phases~\cite{Super-Kamiokande:2011,Super-Kamiokande:2023yqq}, Borexino has demonstrated decade-long stability enabling $>5\sigma$ detection of annual orbital modulation~\cite{Kumaran:2021,BOREXINO:2022wuy,Basilico:2023}, and next-generation liquid xenon detectors (e.g., PandaX-4T, DARWIN) are beginning to observe the solar $^8\mathrm{B}$ ``neutrino fog'' via coherent elastic neutrino-nucleus scattering (CE$\nu$NS)~\cite{Ma:2023Ph,Bo:2024Ph,DeRomeri:2025,DeRomeri:2026JC,Zhuang:2024dm}. These high-statistics, long-baseline measurements---spanning traditional water Cherenkov, liquid scintillator, and emerging dark matter detector platforms---now demand robust analysis frameworks capable of disentangling genuine astrophysical modulations (such as orbital eccentricity or solar rotation harmonics) from detector artifacts, seasonal residuals, and statistical fluctuations.

The Lomb--Scargle (LS) periodogram has long served as the standard tool for detecting periodic signals in unevenly sampled astrophysical time series~\cite{Lomb:1976Ap,Scargle:1982Ap}. However, its application to solar neutrino data reveals several fundamental limitations. The classical LS formulation assumes homoscedastic, Gaussian-distributed noise and strictly sinusoidal signal morphologies. Solar neutrino time series inherently violate these assumptions: they exhibit irregular sampling due to detector maintenance, calibration runs, and phase transitions (e.g., SK-I/II/III/IV hardware upgrades)~\cite{Super-Kamiokande:2005wtt,Super-Kamiokande:2008ecj,Super-Kamiokande:2010tar,Super-Kamiokande:2016yck}, and their statistical uncertainties are heteroscedastic and often asymmetric, particularly in low-statistics bins or near energy thresholds. Moreover, physical modulation mechanisms such as RSFP-driven magnetic suppression, solar rotation harmonics, or complex seasonal residuals after $1/r^2$ Earth--Sun distance correction rarely produce pure sinusoidal waveforms~\cite{BOREXINO:2022wuy}. Traditional periodograms lose significant sensitivity to non-sinusoidal, multi-peaked, or asymmetric modulations, and are highly susceptible to sampling artifacts, data gaps, and outliers. Furthermore, standard variance-based methods may fail to capture anomalous diffusion or non-Gaussian temporal behaviors, such as the L\'evy flight dynamics recently hypothesized in solar neutrino time series~\cite{Haubold:2024ar}. 

Historical analyses illustrate these pitfalls. Early claims of a $\sim 38$-day periodicity in SK-I data, initially attributed to synodic solar core rotation, were shown to diminish to statistical insignificance when the full 22-year dataset was analyzed~\cite{Aharmim:2005Ph,Super-Kamiokande:2023yqq,Pasumarti:2024oei}, highlighting how classical methods can misinterpret fortuitous alignments of sparse data points or unmodeled detector systematics as physical signals. Similarly, standard LS or $\chi^2$-based sinusoidal fits struggle to robustly quantify significance in the presence of correlated noise, non-Gaussian error distributions, or the ``look-elsewhere'' effect across broad frequency grids.

While the Generalized Lomb--Scargle (GLS) method partially mitigates heteroscedasticity by incorporating per-point uncertainties and a floating mean~\cite{Zechmeister:2009AA}, it remains fundamentally constrained by its sinusoidal basis and Gaussian likelihood assumption. To address transient, non-sinusoidal flux depressions that may arise from localized solar activity or instrumental duty-cycle variations, we additionally implement the Box-Fitting Least Squares (BLS) algorithm~\cite{Kovacs:2002}. BLS explicitly models periodic signals as alternating between a baseline and a short-duration depressed state, optimizing sensitivity to box-shaped modulations that sinusoidal or harmonic methods may dilute or mischaracterize~\cite{Ren:2023Ap}. By modeling a fixed dip duration, BLS complements harmonic and phase-folding techniques, ensuring comprehensive coverage of both smooth and abrupt temporal variations in the neutrino flux.

This limitation motivates the adoption of model-agnostic phase-dispersion and information-theoretic frameworks that can extract periodic signals without restrictive parametric assumptions. Classical phase-folding techniques, such as the Lafler--Kinman String Length (LKSL)~\cite{Clarke:2002AA} and Phase Dispersion Minimization (PDM1)~\cite{Stellingwerf:1978Ap}, operate by evaluating the continuity or scatter of data points when folded at trial periods. Rather than fitting a predefined functional form, PDM1 minimizes the within-phase-bin variance relative to the total variance, yielding a statistic that follows an $F$-distribution under the null hypothesis of white noise. This approach is inherently robust to highly non-sinusoidal, multi-peaked, or asymmetric modulation profiles, and it maintains high sensitivity even for sparse datasets with limited phase coverage. The Multi-Harmonic Analysis of Variance (MHAOV)~\cite{Schwarzenberg:1996Ap,Mondrik:2015Ap,Graham:2013MN} extends this framework by explicitly modeling harmonic content through truncated Fourier expansions within an ANOVA framework, offering a flexible bridge between purely non-parametric phase methods and sinusoidal fitting.

More recently, information-theoretic criteria have been adapted for period detection, leveraging measures of statistical dependence that capture both linear and non-linear correlations. The Quadratic Mutual Information (QMI) family---encompassing Euclidean (QMIEU) and Cauchy--Schwarz (QMICS) divergences---quantifies the reduction of uncertainty between phase and flux measurements without requiring explicit probability density function (PDF) estimation~\cite{Huijse:2015ar,Huijse:2018Ap}. By computing information potentials directly from pairwise data distances using kernel-based estimators, QMI methods bypass the strong distributional assumptions of classical techniques. As demonstrated in large-scale astronomical surveys, QMI periodograms exhibit superior resilience to heavy-tailed noise, outliers, and heteroscedastic uncertainties, often recovering true periods at lower signal-to-noise ratios and with fewer observations than second-order methods. The Quadratic Mutual Entropy (QME) variant further stabilizes significance estimation in regimes with irregular sampling and non-stationary backgrounds.

In this work, we present a systematic, multi-method periodogram analysis of the complete 22-year Super-Kamiokande solar $^8\mathrm{B}$ neutrino dataset (1996--2018), comparing nine independent algorithms spanning classical, phase-based, and information-theoretic paradigms. To disentangle potential astrophysical modulations from detector-specific systematics, we employ a hierarchical temporal segmentation strategy, dividing the data into pre- and post-July 2001 epochs and the four distinct operational phases (SK-I--IV)~\cite{Super-Kamiokande:2005wtt,Super-Kamiokande:2008ecj,Super-Kamiokande:2010tar,Super-Kamiokande:2016yck}. This approach effectively isolates true solar-cycle or core-dynamical signatures from instrumental artifacts induced by hardware evolution. By mapping the sensitivity and cross-method consistency of these algorithms across distinct detector configurations, this study establishes a rigorous foundation for interpreting temporal modulations in current and next-generation solar neutrino observatories, including Hyper-Kamiokande~\cite{Hyper-Kamiokande:2018ofw}, DUNE~\cite{Abi:2020ar}, and JUNO~\cite{Abusleme:2022Ch,Stock:2024tmd}.

The paper is organized as follows: Section~\ref{sec:data} details the Super-Kamiokande dataset, preprocessing pipeline, and temporal segmentation strategy. Section~\ref{sec:methods} presents the mathematical formulation and statistical assumptions of the nine periodogram methods, complemented by a Bayesian MCMC sinusoidal fitting framework described in Section~\ref{subsec:bayesian}. Section~\ref{sec:analysis_results} describes the unified analysis pipeline, presents periodogram results across all temporal segments, and evaluates cross-method consistency. Section~\ref{sec:discussion} discusses the physical interpretations, systematic uncertainties, and implications for next-generation detectors. We conclude in Section~\ref{sec:conclusions}. All analysis code and reproducibility instructions are publicly archived.\footnote{\url{https://github.com/renlliang3/sk-neutrino-periodicity}}

\section{Data Description and Preprocessing}
\label{sec:data}

\subsection{Super-Kamiokande Detector and Data Phases}

The Super-Kamiokande (SK) experiment is a 50-kiloton water Cherenkov detector located in the Kamioka mine, Japan, designed to detect solar neutrinos via elastic scattering off atomic electrons ($\nu + e^- \to \nu + e^-$)~\cite{Walter:2008ys}. The standard fiducial volume for solar neutrino analysis is 22.5 kton, providing an average event rate of $\sim$20 interactions per day after accounting for energy thresholds, detection efficiencies, and neutrino oscillations. Over its operational lifetime, SK has undergone four distinct hardware configurations, each introducing specific systematic characteristics that must be accounted for in time-series analyses. The dataset spans from May 31, 1996, to May 30, 2018, accumulating a total of $5804$ live days of solar neutrino observations. The four operational phases are summarized as follows~\cite{Super-Kamiokande:2005wtt,Super-Kamiokande:2008ecj,Super-Kamiokande:2010tar,Super-Kamiokande:2016yck}:

\begin{itemize}
\item \textbf{SK-I} (1996-05-31 to 2001-07-15): 1495.7 live days. Initial operation with 11,146 photomultiplier tubes (PMTs).
\item \textbf{SK-II} (2002-12-10 to 2005-10-06): 791.9 live days. Reduced PMT count (5,182 operational) following the November 2001 accident.
\item \textbf{SK-III} (2006-05-23 to 2008-08-17): 548.5 live days. Full PMT restoration with new hardware and improved calibration.
\item \textbf{SK-IV} (2008-09-15 to 2018-05-30): 2967.7 live days. Upgraded data acquisition (DAQ) system, refined energy calibration, and improved fiducial volume selection.
\end{itemize}

Additionally, we define a \textit{pre/post July 2001} segmentation to isolate the impact of the 2001 PMT failure and subsequent recovery on time-series continuity.

\begin{figure}[tbp]
\centering
\includegraphics[width=15.5cm]{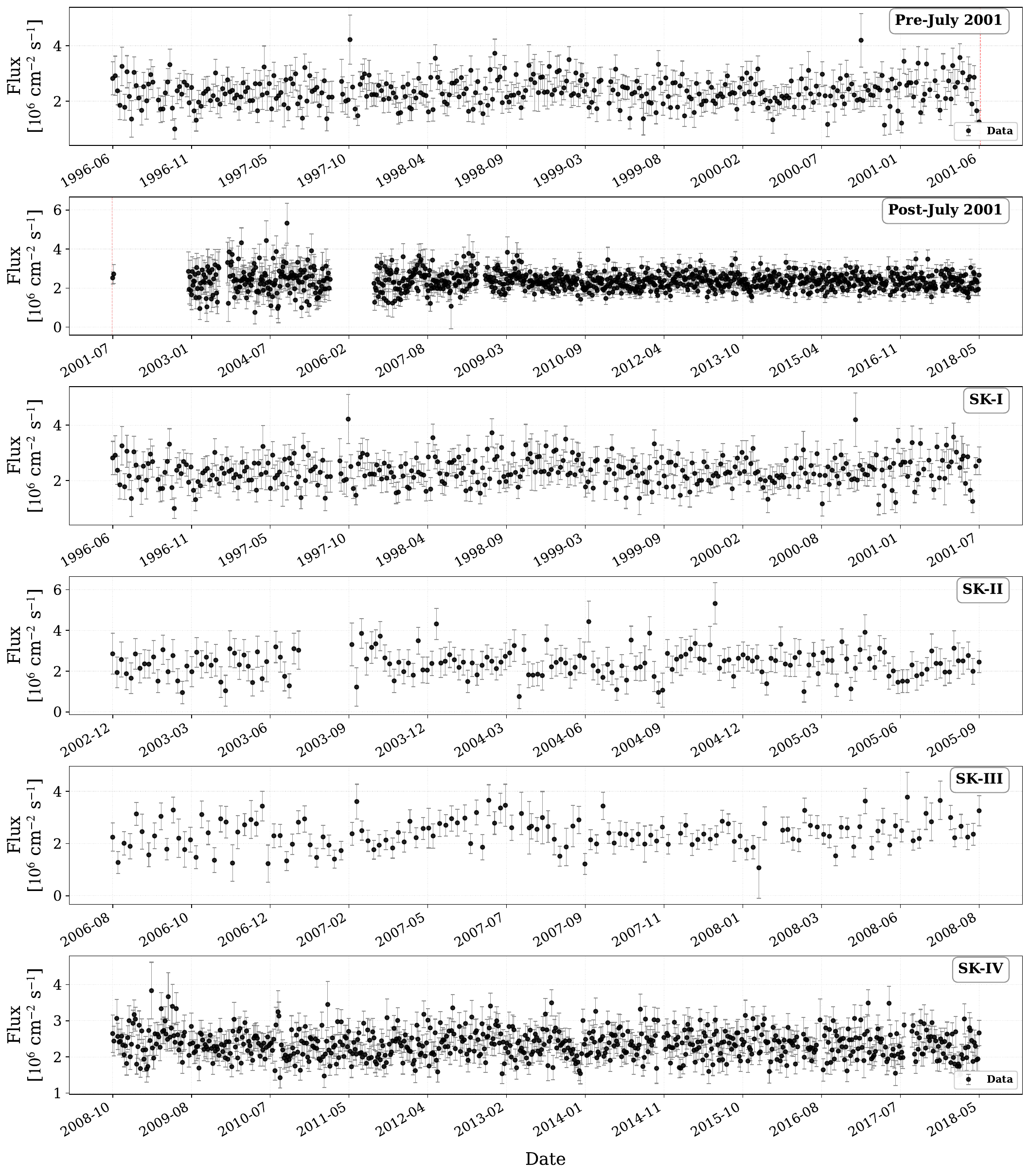}
\caption{Time series of Super-Kamiokande solar $^{8}\mathrm{B}$ neutrino flux measurements binned in nominal 5-day intervals. The six panels display the temporal evolution across hierarchical segments: (top) pre-July 2001 epoch, (second) post-July 2001 epoch, (third to sixth) individual operational phases SK-I, SK-II, SK-III, and SK-IV. Vertical dashed lines mark phase boundaries. Error bars represent heteroscedastic statistical uncertainties derived from extended maximum-likelihood fits. The strong annual modulation driven by Earth's orbital eccentricity is visible prior to distance correction.}
\label{fig:timeseries}
\end{figure}

\subsection{Characteristics of the Time Series Data}
\label{subsec:data_characteristics}
Since the primary innovation of this work lies in the systematic, multi-method periodogram analysis rather than novel data reconstruction, we strictly adopt the data extraction and preprocessing pipeline established by the Super-Kamiokande Collaboration~\cite{Super-Kamiokande:2023yqq} and recently validated by Pasumarti \& Desai~\cite{Pasumarti:2024oei}. The solar neutrino time series used in this study was obtained from the publicly available dataset compiled and released by Pasumarti \& Desai,\footnote{\url{https://github.com/DarkWake9/Project-LP}} which is derived directly from the official Super-Kamiokande 22-year data release.

This dataset consists of flux measurements binned in nominal 5-day intervals, yielding 1,343 data points across the full 22-year span~\cite{Super-Kamiokande:2023yqq,Pasumarti:2024oei}. 
Figure~\ref{fig:timeseries} displays the time series across the hierarchical segments, illustrating the irregular sampling, strong annual modulation, and phase-dependent error structures discussed below. 
Several intrinsic characteristics of the dataset pose challenges for classical spectral analysis, which our multi-method framework is designed to address:
\begin{enumerate}
\item \textbf{Irregular Sampling \& Data Gaps:} The 5-day binning is an approximation; actual bin widths vary due to detector dead time, calibration runs, hardware maintenance, and intentional exclusions~\cite{Super-Kamiokande:2023yqq}. Gaps between phases (SK-I/II and SK-II/III transitions) further disrupt uniform cadence~\cite{Super-Kamiokande:2005wtt,Super-Kamiokande:2008ecj,Super-Kamiokande:2010tar}.
\item \textbf{Strong Seasonal Modulation:} The Earth's elliptical orbit induces a $\sim$7\% annual variation in the solar neutrino flux at the detector~\cite{Super-Kamiokande:2023yqq,Pasumarti:2024oei}, peaking in early January (perihelion). This dominant seasonal signal must be precisely modeled and removed to avoid aliasing into lower-frequency bands.
\item \textbf{Heteroscedastic \& Asymmetric Errors:} Flux uncertainties in each bin are derived from the curvature of the extended maximum likelihood profile and are inherently asymmetric. The statistical variance scales with the inverse of the event count, while phase-dependent systematic uncertainties vary across SK-I--IV~\cite{Super-Kamiokande:2005wtt,Super-Kamiokande:2008ecj,Super-Kamiokande:2010tar,Super-Kamiokande:2016yck}.
\item \textbf{Dead-Time \& Trigger Evolution:} Changes in trigger logic and spallation veto efficiencies introduce time-dependent livetime corrections. The effective exposure per bin is explicitly accounted for in the flux extraction likelihood.
\item \textbf{Energy Threshold Evolution:} The analysis threshold dropped from 6.49 MeV (SK-II) to 3.49 MeV (late SK-IV)~\cite{Super-Kamiokande:2005wtt,Super-Kamiokande:2008ecj,Super-Kamiokande:2010tar,Super-Kamiokande:2016yck}, altering the relative contribution of the $^8\mathrm{B}$ spectrum tail and changing the signal-to-background ratio.
\end{enumerate}

\subsection{Preprocessing Pipeline}
\label{subsec:preprocessing}
To prepare the time series for periodicity searches, we implement a standardized preprocessing pipeline that isolates intrinsic flux variations from orbital and instrumental artifacts, strictly following the methodology detailed in Refs.~\cite{Super-Kamiokande:2023yqq,Pasumarti:2024oei}. 

The primary correction addresses the Keplerian annual modulation. Following the procedure established in~\cite{Pasumarti:2024oei}, raw 5-day flux values are multiplied by the squared Earth--Sun distance $D(t)^2$, normalized to 1 astronomical unit (AU). The distance $D(t)$ is computed using the \texttt{astropy} solar system ephemeris, then bin-averaged over each 5-day interval to match the observational cadence. This transformation flattens the $\sim$7\% seasonal baseline, preventing the orbital eccentricity from aliasing into lower-frequency bands and ensuring that any residual modulations reflect genuine solar or detector-driven phenomena~\cite{BOREXINO:2022wuy}.
\begin{equation}
F_{\rm corr}(t) = F_{\rm obs}(t) \times \left[ \frac{D(t)}{\langle D \rangle} \right]^2,
\label{eq:d2_correction}
\end{equation}
where $D(t)$ is the instantaneous Sun-Earth distance and $\langle D \rangle$ is the annual mean. Uncertainties are propagated accordingly. 

Given the asymmetric statistical uncertainties $(\sigma_-, \sigma_+)$ reported in the original dataset, we symmetrize the errors as $\sigma = (\sigma_- + \sigma_+)/2$. This symmetrization approach is explicitly adopted from Pasumarti \& Desai~\cite{Pasumarti:2024oei} to maintain compatibility with standard periodogram implementations (which typically assume Gaussian errors) while preserving the effective information content. Systematic uncertainties from energy calibration, fiducial volume reconstruction, and spallation background modeling are conservatively added in quadrature to the statistical errors. This approach ensures that phase-dependent instrumental systematics do not artificially inflate spectral power or generate spurious peaks \cite{Super-Kamiokande:2023yqq,Huijse:2018Ap}. All flux measurements are expressed in units of $10^6 \, \text{cm}^{-2} \, \text{s}^{-1}$ under the assumption of no neutrino oscillations, facilitating direct comparison with Standard Solar Model predictions. Time stamps for each bin are assigned as the livetime-weighted mean of the bin boundaries, correcting for non-uniform data acquisition within each interval and preserving accurate phase relationships for epoch-folding algorithms~\cite{Super-Kamiokande:2023yqq}.

\subsection{Data Segmentation Strategy}

Given the heterogeneous nature of the 22-year dataset, we employ a hierarchical temporal segmentation strategy to disentangle astrophysical modulations from detector-specific systematics and solar-cycle dependencies. The analysis is first performed on the full 22-year baseline, which maximizes statistical power and probes long-term stability across nearly two complete solar activity cycles (Cycles 23 and 24). This global view serves as the primary benchmark for cross-method periodogram comparison and establishes the overall sensitivity floor for sub-percent modulation searches~\cite{Super-Kamiokande:2023yqq,Pasumarti:2024oei}.

To isolate the impact of major detector upgrades, we split the dataset at July 2001, separating the SK-I era (pre-accident, 40\% PMT coverage) from all subsequent reconstruction phases. This boundary coincides with the transition from Solar Cycle 23's rising phase to its maximum, as well as the fundamental hardware shift to FRP-shielded PMTs and modified reconstruction algorithms. Comparing Pre/Post 2001 segments allows us to test whether candidate periodicities persist across fundamentally different detector configurations or are artifacts of early-phase reconstruction biases or the historically debated $\sim$38-day statistical fluctuation~\cite{Super-Kamiokande:2023yqq,Pasumarti:2024oei}.

Finally, we conduct independent periodogram analyses for each operational phase (SK-I through SK-IV). This phase-by-phase segmentation explicitly controls for phase-specific systematics: SK-II's reduced light yield and higher threshold, SK-III's threshold reduction and improved vertex fitting, and SK-IV's QBEE electronics upgrades, refined spallation tagging via neutron clustering, and dynamic trigger efficiency corrections~\cite{Super-Kamiokande:2023yqq,Pasumarti:2024oei}. A physically meaningful periodic signal should manifest consistently across phases (allowing for expected statistical fluctuations and solar-cycle amplitude modulation), whereas detector-driven artifacts (e.g., seasonal water transparency drifts, PMT gain evolution, or trigger-threshold aliases) will be confined to specific segments. This multi-tiered segmentation framework ensures that any detected periodicity is robust against hardware evolution, systematic drift, and solar-cycle phase, providing a rigorous foundation for the statistical assessment presented in subsequent sections.

\section{Methodology}
\label{sec:methods}

\subsection{Classical and Generalized Lomb-Scargle Methods}

\subsubsection*{Lomb--Scargle Periodogram (\texttt{LS\_Standard})}

The classical Lomb--Scargle (LS) periodogram~\citep{Lomb:1976Ap,Scargle:1982Ap} searches for periodic sinusoidal components by fitting the model $y(t) = a \cos \omega t + b \sin \omega t$ via least squares. It assumes homoscedastic, Gaussian-distributed measurement errors and a zero-mean data series. The normalized LS power at frequency $\omega$ is computed as~\citep{Leroy:2012AA,VanderPlas:2015Ap,VanderPlas:2018Ap}:

\begin{equation}
p_{\rm LS}(\omega) = \frac{1}{2\sigma^2} \left[ \frac{\left[\sum_{i=1}^N y_i \cos \omega(t_i - \tau)\right]^2}{\sum_{i=1}^N \cos^2 \omega(t_i - \tau)} + \frac{\left[\sum_{i=1}^N y_i \sin \omega(t_i - \tau)\right]^2}{\sum_{i=1}^N \sin^2 \omega(t_i - \tau)} \right],
\end{equation}
where $\sigma^2 = \sum_i (y_i - \bar{y})^2 / (N-1)$ is the sample variance, $\bar{y}$ is the arithmetic mean, and the time-shift parameter $\tau$ is defined to orthogonalize the sine and cosine terms~\citep{Leroy:2012AA,VanderPlas:2015Ap,VanderPlas:2018Ap}:

\begin{equation}
\tan(2\omega\tau) = \frac{\sum_{i=1}^N \sin 2\omega t_i}{\sum_{i=1}^N \cos 2\omega t_i}.
\end{equation}

While computationally efficient and analytically tractable for significance estimation, the standard LS formulation is highly sensitive to mean offsets and assumes uniform error weighting, making it suboptimal for neutrino flux bins with asymmetric or heteroscedastic uncertainties.

\subsubsection*{Generalized Lomb--Scargle (\texttt{GLS})}

To address the limitations of the classical LS, the Generalized Lomb--Scargle (GLS) periodogram~\citep{Zechmeister:2009AA} incorporates per-point measurement uncertainties and a floating mean offset $c$. The model becomes $y(t) = a \cos \omega t + b \sin \omega t + c$, and the fit minimizes the weighted $\chi^2$ statistic~\citep{Zechmeister:2009AA}:

\begin{equation}
\chi^2(\omega) = \sum_{i=1}^N w_i \left[ y_i - (a \cos \omega t_i + b \sin \omega t_i + c) \right]^2,
\end{equation}
where $w_i = 1/\sigma_i^2$ are the inverse-variance weights. The GLS power is normalized to the $\chi^2$ of the weighted mean model ($\chi_0^2$) and bounded to $[0, 1]$~\citep{Zechmeister:2009AA}:

\begin{equation}
p_{\rm GLS}(\omega) = \frac{\chi_0^2 - \chi^2(\omega)}{\chi_0^2} = \frac{1}{YY_\tau} \left( \frac{YC_\tau^2}{CC_\tau} + \frac{YS_\tau^2}{SS_\tau} \right),
\end{equation}
with the weighted sums defined relative to the weighted mean $\bar{y}_w = \sum w_i y_i / \sum w_i$ and a frequency-dependent time shift $\tau$. The GLS reduces exactly to the classical LS under equal weights and zero-mean data, but provides superior peak sensitivity and robustness against data gaps and heteroscedastic noise in solar neutrino time series.

\subsection{Box-Fitting Least Squares Method}

\subsubsection*{Box Least Squares Periodogram (\texttt{BLS})}

The Box Least Squares (BLS) algorithm~\cite{Kovacs:2002,Hartman:2016AC} is specifically designed to detect periodic signals characterized by brief, box-shaped dips---a morphology distinct from sinusoidal or smoothly varying waveforms. This makes BLS particularly suitable for identifying transit-like modulations or short-duration flux depressions in time series data, such as those potentially arising from solar rotational modulation of neutrino production regions or detector-related systematic effects.

The BLS model assumes a strictly periodic signal with period $P_0$ that alternates between two discrete levels~\cite{Kovacs:2002}: a baseline $H$ and a depressed level $L$ sustained for a fractional duration $q \equiv \Delta t / P_0 \ll 1$. The epoch of the dip center is denoted $t_0$. Under the constraint of zero-mean signal, the levels satisfy $H = -L q / (1-q)$, reducing the free parameters to four: $\{P_0, q, L, t_0\}$.

For a trial period $P$, the observed time series $\{t_i, y_i, \sigma_i\}$ is phase-folded via $\phi_i = (t_i / P) \bmod 1$ and sorted by phase. The data are optionally binned into $m$ phase bins to improve computational efficiency. For any candidate dip window defined by bin indices $[i_1, i_2]$, the weighted residual sum of squares is minimized with respect to the two levels~\cite{Kovacs:2002}:

\begin{equation}
D(i_1, i_2) = \sum_{i \notin [i_1,i_2]} w_i (y_i - \hat{H})^2 + \sum_{i \in [i_1,i_2]} w_i (y_i - \hat{L})^2,
\end{equation}
where $w_i = \sigma_i^{-2} / \sum_j \sigma_j^{-2}$ are normalized inverse-variance weights. Analytic minimization yields the level estimators~\cite{Kovacs:2002}:

\begin{equation}
\hat{L} = \frac{s}{r}, \qquad \hat{H} = -\frac{s}{1-r},
\end{equation}
with the weighted sums defined as~\cite{Kovacs:2002}:

\begin{equation}
s(i_1,i_2) = \sum_{i=i_1}^{i_2} w_i y_i, \qquad r(i_1,i_2) = \sum_{i=i_1}^{i_2} w_i.
\end{equation}

Substituting these estimates, the minimized residual becomes~\cite{Kovacs:2002}:

\begin{equation}
D_{\rm min}(i_1,i_2) = \sum_{i=1}^N w_i y_i^2 - \frac{s^2(i_1,i_2)}{r(i_1,i_2)[1-r(i_1,i_2)]}.
\end{equation}

Since the first term is independent of the trial period and window, the quality of fit is characterized by the \textit{Signal Residue} (SR) statistic~\cite{Kovacs:2002,Hartman:2016AC}:

\begin{equation}
{\rm SR}(P) = \max_{i_1,i_2} \left\{ \frac{|s(i_1,i_2)|}{\sqrt{r(i_1,i_2)[1-r(i_1,i_2)]}} \right\},
\label{eq:bls_sr}
\end{equation}
where the maximization is performed over all admissible windows satisfying $q_{\rm min} \leq (i_2-i_1)/m \leq q_{\rm max}$. The BLS periodogram is defined as ${\rm SR}(P)$ evaluated over a grid of trial periods.

The detection significance of a BLS peak is governed by the \textit{effective signal-to-noise ratio}~\cite{Kovacs:2002}:

\begin{equation}
\alpha \equiv \frac{\delta}{\sigma} \sqrt{N q},
\label{eq:bls_snr}
\end{equation}
where $\delta = H - L$ is the dip depth, $\sigma$ is the typical photometric uncertainty, $N$ is the number of data points, and $q$ is the fractional dip duration. Numerical simulations demonstrate that $\alpha \gtrsim 6$ is required for reliable detection at the $3\sigma$ level, with the Signal Detection Efficiency (SDE) rising sharply in the range $\alpha = 6$--$13$ ~\cite{Kovacs:2002}.

In our implementation, we adopt a fixed dip duration of 10 days, phase binning $m=100$, a linearly spaced grid of 1000 trial periods covering the range from 18.26 days to $\min(0.8 \times T_{\rm span}, 5 \text{ yr})$, and significance assessment via 100 bootstrap realizations combined with extreme value theory. BLS offers distinct advantages for neutrino time-series analysis: optimal sensitivity to short-duration, non-sinusoidal dips that may be diluted by sinusoidal-fitting methods; explicit modeling of the dip duty cycle; and robustness to heteroscedastic uncertainties through weighted least squares.

Classical-LS, GLS and BLS methods are implemented using the \texttt{Astropy} Python package.\footnote{\url{https://github.com/astropy/astropy}}

\subsection{Phase-Dispersion and Harmonic Methods}

\subsubsection*{Lafler--Kinman String Length (\texttt{LKSL})}

The LKSL method~\citep{Clarke:2002AA} evaluates the smoothness of a phase-folded light curve without assuming a functional form. For a trial period $P$, observation times are folded to phases $\phi_i = (t_i / P) \bmod 1$. The data points are sorted by phase, and the string length statistic is computed as the sum of squared differences between adjacent points~\citep{Clarke:2002AA}:

\begin{equation}
S(P) = \sum_{i=1}^{N-1} (y_{i+1} - y_i)^2.
\end{equation}

A true periodic signal minimizes $S(P)$, as correctly folded points align along a continuous trajectory. The normalized periodogram is $p_{\rm LKSL}(P) = 1 - S(P)/S_{\rm max}$, where $S_{\rm max}$ corresponds to a randomized phase distribution. LKSL is highly sensitive to non-sinusoidal, steep-edged, or multi-peak modulations but lacks an intrinsic analytical false-alarm probability distribution.

\subsubsection*{Phase Dispersion Minimization (\texttt{PDM1})}

PDM1 quantifies the scatter of data points around a mean phase curve~\citep{Stellingwerf:1978Ap}. The phase interval $[0, 1)$ is divided into $M$ bins. For each trial period, the within-bin variance $\sigma_j^2$ and the total variance $\sigma^2$ are computed. The PDM statistic is the ratio~\citep{Stellingwerf:1978Ap}:

\begin{equation}
\Theta(P) = \frac{s^2}{\sigma^2}, \quad \text{where} \quad s^2 = \frac{1}{N-M} \sum_{j=1}^M (n_j - 1) \sigma_j^2,
\end{equation}
with $n_j$ the number of points in bin $j$. Under the null hypothesis of white noise, $\Theta \approx 1$; a true period yields $\Theta \ll 1$. The periodogram is typically inverted as $p_{\rm PDM}(P) = 1 - \Theta(P)$. PDM1 is distribution-free, robust to asymmetric waveforms, and follows an $F$-distribution under Gaussian noise assumptions, enabling analytic significance testing.

\subsubsection*{Multi-Harmonic Analysis of Variance (\texttt{MHAOV})}

MHAOV extends the ANOVA framework to non-sinusoidal signals by fitting a truncated Fourier series with $K$ harmonics~\citep{Schwarzenberg:1996Ap,Mondrik:2015Ap,Graham:2013MN,Hartman:2016AC}:

\begin{equation}
y(t) = A_0 + \sum_{k=1}^K \left[ A_k \cos(k\omega t) + B_k \sin(k\omega t) \right].
\end{equation}

The method partitions the total sum of squares into model variance ($SS_{\rm model}$) and residual variance ($SS_{\rm res}$). The test statistic follows an $F$-distribution with $(2K, N-2K-1)$ degrees of freedom~\citep{Schwarzenberg:1996Ap,Mondrik:2015Ap}:

\begin{equation}
F(P) = \frac{SS_{\rm model} / 2K}{SS_{\rm res} / (N - 2K - 1)}.
\end{equation}

The periodogram $p_{\rm MHAOV}(P)$ is proportional to $F(P)$. MHAOV effectively bridges parametric harmonic fitting and non-parametric phase binning, offering high sensitivity to complex, quasi-periodic modulations while maintaining rigorous statistical interpretability.

\subsection{Information-Theoretic Periodogram Methods}

Information-theoretic periodograms bypass explicit waveform modeling by quantifying the statistical dependence between folded phases $\Phi$ and observed fluxes $F$. We adopt the Quadratic Mutual Information (QMI) framework ~\citep{Huijse:2015ar,Huijse:2018Ap}, which operates directly on data samples via kernel-based Information Potentials (IPs), avoiding explicit probability density function (PDF) estimation.

\subsubsection*{Information Potential Estimation}

Let $\{\phi_i, f_i\}_{i=1}^N$ denote the phase-folded time series for a trial period. We define two kernel functions: a Gaussian kernel for fluxes $G_h(x) = \frac{1}{\sqrt{2\pi}h} \exp(-x^2/2h^2)$ and a Wrapped Cauchy kernel for periodic phases $WC_h(\Delta\phi) = \frac{1}{2\pi} \frac{1 - e^{-2h}}{1 + e^{-2h} - 2e^{-h}\cos(2\pi\Delta\phi)}$. The joint and marginal IPs are computed as~\citep{Huijse:2015ar,Huijse:2018Ap}:

\begin{align}
IP_F &= \frac{1}{N^2} \sum_{i,j} G_{\sqrt{2}h_f}(f_i - f_j), \quad
IP_\Phi = \frac{1}{N^2} \sum_{i,j} WC_{2h_\phi}(\phi_i - \phi_j), \nonumber\\
IP_{F,\Phi} &= \frac{1}{N^2} \sum_{i,j} G_{\sqrt{2}h_f}(f_i - f_j) \, WC_{2h_\phi}(\phi_i - \phi_j), \nonumber\\
IP_{F \times \Phi} &= \frac{1}{N} \sum_{i=1}^N \left[ \frac{1}{N} \sum_{j=1}^N G_{\sqrt{2}h_f}(f_i - f_j) \right] \left[ \frac{1}{N} \sum_{j=1}^N WC_{2h_\phi}(\phi_i - \phi_j) \right].
\end{align}

The bandwidth $h_f$ is selected via a weighted Silverman's rule-of-thumb to account for heteroscedastic flux errors, while $h_\phi = 1$ is empirically optimal for phase variables~\citep{Huijse:2015ar,Huijse:2018Ap}.

\subsubsection*{QMI Euclidean (\texttt{QMIEU})}

The Euclidean QMI measures the $L_2$ divergence between the joint PDF and the product of marginals~\citep{Huijse:2015ar,Huijse:2018Ap}:

\begin{equation}
QMI_{\rm EU}(\Phi, F) = IP_{F,\Phi} - 2 IP_{F \times \Phi} + IP_F IP_\Phi.
\end{equation}

Correctly folded data maximize the statistical distance between joint and independent structures, yielding a sharp periodogram peak.

\subsubsection*{QMI Cauchy--Schwarz (\texttt{QMICS})}

The Cauchy--Schwarz variant replaces the $L_2$ norm with a logarithmic divergence, enhancing robustness to outliers and heavy-tailed noise~\citep{Huijse:2015ar,Huijse:2018Ap}:

\begin{equation}
QMI_{\rm CS}(\Phi, F) = \log(IP_{F,\Phi}) - 2\log(IP_{F \times \Phi}) + \log(IP_F) + \log(IP_\Phi).
\end{equation}

Both $QMI_{\rm EU}$ and $QMI_{\rm CS}$ are non-negative, vanish under statistical independence, and are invariant to monotonic flux transformations.

\subsubsection*{Quadratic Mutual Entropy (\texttt{QME})}

QME is formulated via R\'enyi entropy of order 2, $H_2(X) = -\log(IP_X)$. The mutual entropy quantifies the reduction in uncertainty of $F$ given $\Phi$~\citep{Huijse:2015ar,Huijse:2018Ap}:

\begin{equation}
QME(\Phi, F) = H_2(F) + H_2(\Phi) - H_2(\Phi, F) = \log(IP_F) + \log(IP_\Phi) - \log(IP_{F,\Phi}).
\end{equation}

Maximizing $QME$ over trial periods is mathematically equivalent to minimizing the joint IP $IP_{F,\Phi}$, emphasizing compact phase clustering. Information-theoretic methods require no sinusoidal assumption, naturally handle heteroscedastic and non-Gaussian errors, and have been shown to outperform second-order statistics in sparse, low-SNR regimes~\citep{Huijse:2015ar,Huijse:2018Ap}.

All P4J methods are implemented using the \texttt{P4J} Python package.\footnote{\url{https://github.com/phuijse/P4J}}

\subsection{Statistical Significance}
\label{subsec:significance}

To rigorously assess peak significance, we employ method-specific approaches tailored to their statistical properties. For the LS and GLS methods, we utilize the analytical approximation proposed by Baluev (2008) to compute the False Alarm Probability (FAP)~\citep{Baluev:2008MN,Baluev:2009MN}, which is computationally efficient and robust for these specific periodograms. 

For the remaining methods (BLS and the P4J suite), we employ a non-parametric block bootstrap resampling framework combined with extreme value theory (EVT). Specifically, we generate 100 synthetic light curves by resampling 30-day blocks of the observed data while preserving the original observation times $t_i$ and uncertainties $\sigma_i$. The empirical FAP for a peak power $P_{\rm max}$ is:

\begin{equation}
{\rm FAP}(P_{\rm max}) = \frac{1}{N_{\rm boot}} \sum_{k=1}^{N_{\rm boot}} \mathbb{I}\left( \max_\omega P_k(\omega) \geq P_{\rm max} \right),
\end{equation}
where $\mathbb{I}(\cdot)$ is the indicator function. This accounts for the ``look-elsewhere'' effect and window-function aliasing inherent in irregular sampling. To estimate FAPs below $1/N_{\rm boot}$, we fit a Generalized Extreme Value (GEV) distribution to the bootstrap maxima and compute the tail probability.

\subsection{Bayesian Sinusoidal Fitting and MCMC Validation}
\label{subsec:bayesian}

To validate the periodogram detections and rigorously quantify parameter uncertainties, we implement a Bayesian inference pipeline that strictly adheres to the time-binned formalism established by the Super-Kamiokande collaboration~\cite{Super-Kamiokande:2023yqq}. Unlike instantaneous point-value models, the solar neutrino flux is measured over finite five-day intervals. Following Eq.~(2) of Ref.~\cite{Super-Kamiokande:2023yqq}, the expected flux in the $r$-th time bin is given by the bin-averaged modulation:

\begin{equation}
g_r(\omega; A, B) = g_0 + \frac{1}{t_{r,f} - t_{r,i}} \int_{t_{r,i}}^{t_{r,f}} \! dt \, \bigl( A \cos \omega t + B \sin \omega t \bigr),
\label{eq:prl_bin_avg_integral}
\end{equation}
where $t_{r,i}$ and $t_{r,f}$ denote the initial and final times of the $r$-th interval, $\omega = 2\pi f$ is the angular frequency (with $f$ in yr$^{-1}$), and $A$ and $B$ are the free cosine and sine amplitudes. The baseline $g_0$ corresponds to the mean flux level. In our implementation, we treat $g_0$ as a free offset parameter $c$ and add a linear drift term $s \cdot t_{c,r}$ to account for long-term instrumental trends. Analytically evaluating the integral yields the computationally efficient form used in the MCMC sampler:

\begin{equation}
g_r(\omega; A, B, c, s) = c + s \cdot t_{c,r} + \bigl( A \cos \omega t_{c,r} + B \sin \omega t_{c,r} \bigr) \cdot \mathrm{sinc}\!\left( \frac{\omega \Delta t_r}{2} \right),
\label{eq:prl_bin_avg_analytic}
\end{equation}
where $t_{c,r} = (t_{r,i} + t_{r,f})/2$ is the livetime-weighted bin center, $\Delta t_r = t_{r,f} - t_{r,i}$ is the bin width, and $\mathrm{sinc}(x) \equiv \sin(x)/x$. This attenuation factor exactly captures the high-frequency damping induced by finite bin averaging, as derived from the integral in Eq.~\eqref{eq:prl_bin_avg_integral}.

\paragraph{Asymmetric Profile Likelihood.}

The statistical uncertainties of the five-day flux measurements are asymmetric, extracted from the curvature of the extended maximum-likelihood profile~\cite{Super-Kamiokande:2023yqq}. We therefore adopt the exact asymmetric log-likelihood formulation defined in the original analysis:

\begin{equation}
-\ln \mathcal{L}(A, B, f, c, s) = \frac{1}{2} \sum_{r=1}^{N} \left[ \frac{D_r - g_r(\omega; A, B, c, s)}{\sigma_r} \right]^2,
\label{eq:prl_asym_likelihood}
\end{equation}
where $D_r$ is the observed distance-corrected flux in bin $r$. The effective uncertainty $\sigma_r$ is selected conditionally based on the sign of the residual, exactly as specified in Ref.~\cite{Super-Kamiokande:2023yqq}:

\begin{equation}
\sigma_r = \begin{cases}
\sigma_{+,r} & \text{if } D_r < g_r \quad (\text{model} > \text{data}), \\
\sigma_{-,r} & \text{if } D_r > g_r \quad (\text{model} < \text{data}),
\end{cases}
\end{equation}
with $\sigma_{+,r}$ and $\sigma_{-,r}$ denoting the reported upward and downward statistical errors. This piecewise Gaussian likelihood rigorously propagates the asymmetric nature of the SK flux extraction without artificial symmetrization.

\paragraph{Bayesian Inference and Model Comparison.}

We sample the posterior distribution $\mathcal{P}(\theta \mid \mathcal{D}) \propto \mathcal{L}(\theta) \, \pi(\theta)$ using the affine-invariant MCMC ensemble sampler \texttt{emcee}~\citep{Foreman:2013PA}. The parameter vector is $\theta = \{A, B, f, c, s\}$. Uniform priors are applied within data-driven bounds: $A, B \sim \mathcal{U}(-5\sigma_F, 5\sigma_F)$, $f \sim \mathcal{U}(0.6 f_{\rm init}, 1.4 f_{\rm init})$, $c \sim \mathcal{U}(\bar{F} \pm 3\sigma_F)$, and $s \sim \mathcal{U}(-10^{-3}, 10^{-3})$, where $f_{\rm init}$ is the candidate frequency from the periodogram peak and $\sigma_F$ is the flux standard deviation. We employ 128 walkers, 30000 production steps, and discard the first 200 steps as burn-in. Convergence is verified via integrated autocorrelation times and Gelman--Rubin statistics.

We adopt three complementary criteria to evaluate the fitted model:

\begin{itemize}
\item \textbf{Chi-squared} ($\chi^2$) measures the absolute fit~\cite{Andrae:2010ar}:
\begin{equation}
\chi^2 = \sum_{i=1}^{N} \frac{(F_i - \hat{F}(t_i;\theta))^2}{\sigma_i^2}.
\end{equation}

\item \textbf{Bayesian Information Criterion} (BIC) penalizes model complexity~\cite{Konishi:2008}:
\begin{equation}
\mathrm{BIC} = \chi^2 + k\ln N,
\end{equation}
where $k=5$ is the number of free parameters (the sinusoidal model) and $N$ is the number of data points. A lower BIC indicates a more parsimonious model.

\item \textbf{Bayes Factor} (BF~\cite{Speagle:2020MN}) compares the sinusoidal model against a null model consisting only of a constant plus linear trend ($A=0$). Using the BIC approximation~\citep{Faulkenberry:2018ar}, the Bayes factor is:
\begin{equation}
\mathrm{BF} \approx \exp\left(\frac{\mathrm{BIC}_{\mathrm{null}} - \mathrm{BIC}_{\mathrm{model}}}{2}\right).
\end{equation}
The logarithmic Bayes factor is defined as the ratio of the evidence of these two competing models~\citep{Faulkenberry:2018ar}:
\begin{equation}
\ln \mathrm{B} \approx \frac{\mathrm{BIC}_{\mathrm{null}} - \mathrm{BIC}_{\mathrm{model}}}{2}.
\end{equation}

The criterion for interpreting the Bayes factor is given by Jeffreys' scale \citep{Jeffreys:1961}, which classifies the strength of evidence based on the natural logarithm of the Bayes factor, $\ln \mathrm{B}$. In favor of the sinusoidal model, the scale is categorized as follows~\cite{Alfradique:2025ar}:
\begin{itemize}
\item \textbf{Strong evidence}: if $\ln \mathrm{B} > 5$;
\item \textbf{Moderate evidence}: if $2 < \ln \mathrm{B} \le 5$;
\item \textbf{Weak evidence}: if $0 < \ln \mathrm{B} \le 2$;
\item \textbf{No evidence} (models are equally supported): if $\ln \mathrm{B} \le 0$.
\end{itemize}
If $\ln \mathrm{B}$ assumes negative values, the null model is better supported than the sinusoidal model, classified symmetrically as strong ($\ln \mathrm{B} < -5$), moderate ($-5 \le \ln \mathrm{B} < -2$), or weak ($-2 \le \ln \mathrm{B} < 0$) evidence in favor of the null model.
\end{itemize}

The posterior probability that the amplitude $A$ is positive, $P(A>0)$, is computed directly from the MCMC samples. Phase-folded light curves are generated for visual coherence checks, with data binned into 25 phase intervals to suppress statistical noise while preserving morphological features.

\section{Analysis and Results}
\label{sec:analysis_results}

\subsection{Algorithm Implementation and Computational Details}

Before presenting the periodogram results, we detail the numerical implementation of the nine algorithms as executed in our analysis pipeline. All calculations were performed in Python 3.10 using the \texttt{astropy.timeseries} (for LS/GLS/BLS) and \texttt{P4J} (for LKSL, PDM1, MHAOV, QME, QMICS, QMIEU) libraries. The key computational parameters are:

\begin{itemize}
\item \textbf{Frequency grid:} Uniformly spaced from $0.01$ yr$^{-1}$ to $20$ yr$^{-1}$ with a step size determined by $\Delta f = 1/(T \cdot N_{\rm samp})$, where $T$ is the time span of the segment (in years) and $N_{\rm samp}=5$ samples per independent frequency peak.
\item \textbf{BLS configuration:} Trial periods linearly spaced between $18.26$ d and $\min(0.8 \times T_{\rm span}, 5 \text{ yr})$, comprising 1000 points; the dip duration is fixed to 10 days; phase binning $m=100$; oversampling factor 1.
\item \textbf{P4J settings:} For all phase-dispersion and information-theoretic methods, the frequency grid is first evaluated at the same resolution as LS/GLS, then a local refinement around the five highest peaks is performed with 10$\times$ higher resolution. Kernel bandwidths: $h_f = 0.5 \cdot \mathrm{std}(F) \cdot N^{-1/6}$ (weighted by errors), $h_\phi = 1.0$ (wrapped Cauchy).
\item \textbf{Block bootstrap:} To preserve temporal correlations in the noise, we use a block length of 30 days (six 5-day bins). For each of the $N_{\rm boot}=100$ realizations, we resample blocks with replacement until the original time span is covered, then shuffle the order of blocks. This method accounts for potential short-term detector systematics.
\item \textbf{GEV fitting:} The maximum power/criterion from each bootstrap realization is recorded. A GEV distribution (type I, II, or III) is fitted to these maxima using scipy.stats.genextreme. The 95\% significance threshold is the GEV quantile corresponding to $1-\alpha$ (or $\alpha$ for lower-is-better statistics).
\item \textbf{Bayesian MCMC:} For each segment and candidate period (taken from the GLS/LS peak), we run \texttt{emcee} with 128 walkers, 30000 steps (burn-in 200). Convergence is checked by the autocorrelation time. The resulting Bayes factor and posterior probability of $A>0$ are reported.
\end{itemize}

All computations were executed on a workstation with Intel Xeon Gold 6230 CPU (2.1 GHz, 20 cores) and 128 GB RAM. The full analysis of all segments (9 methods $\times$ 6 segments $\times$ 2 flux types) took approximately 28 hours, with bootstrap resampling dominating the runtime.

\subsection{Full Dataset (22 yr) Results}
\label{subsec:Full_Results}

Application of the nine periodogram methods to the full 22-year time series reveals a clear hierarchy of sensitivity across algorithms. As summarized in Table~\ref{tab:period_temporal_raw_mod} and the comprehensive Bayesian parameter estimates in Appendix Table~\ref{tab:combined_bestfit_params}, the classical LS method fails to identify any significant period in either raw or modified flux (FAP = 0.99 in both cases), highlighting its severe vulnerability to heteroscedastic uncertainties and irregular sampling. In stark contrast, the GLS method, which incorporates per-point uncertainties, successfully recovers the dominant period at $P = 0.9382$ yr ($f = 1.066$ yr$^{-1}$) in raw flux with a FAP $< 0.001$. This peak corresponds to the residual annual modulation. For the modified (distance-corrected) flux, the LS method again fails (FAP = 0.99), while the GLS method identifies a dominant peak at $P = 0.1064$ yr ($f = 9.395$ yr$^{-1}$) with FAP $< 0.001$, corresponding to approximately 38.8 days. Information-theoretic methods (QME, QMICS, QMIEU) and phase-dispersion techniques (PDM1, MHAOV) do not yield statistically significant detections in the full dataset (all FAPs $> 0.1$), indicating that the dominant periodic structure is well captured by sinusoidal models. Similarly, the BLS method does not identify any significant periodicity (FAPs $> 0.9$). 

A comprehensive visual comparison of all nine algorithms applied to both raw and modified flux measurements is provided in Appendix Figure~\ref{fig:full_periodogram}. Notably, the periodograms for the information-theoretic algorithms (QME, QMICS, and QMIEU) appear to fail or exhibit severe degeneracies in this specific dataset, yielding flat or boundary-peaked power spectra without statistically significant features. This suggests that the kernel-based information potentials may struggle with the specific noise properties and sparse sampling of the 22-year neutrino time series compared to the variance-based GLS approach.

\begin{figure}[tbp]
\centering
\includegraphics[width=0.9\textwidth]{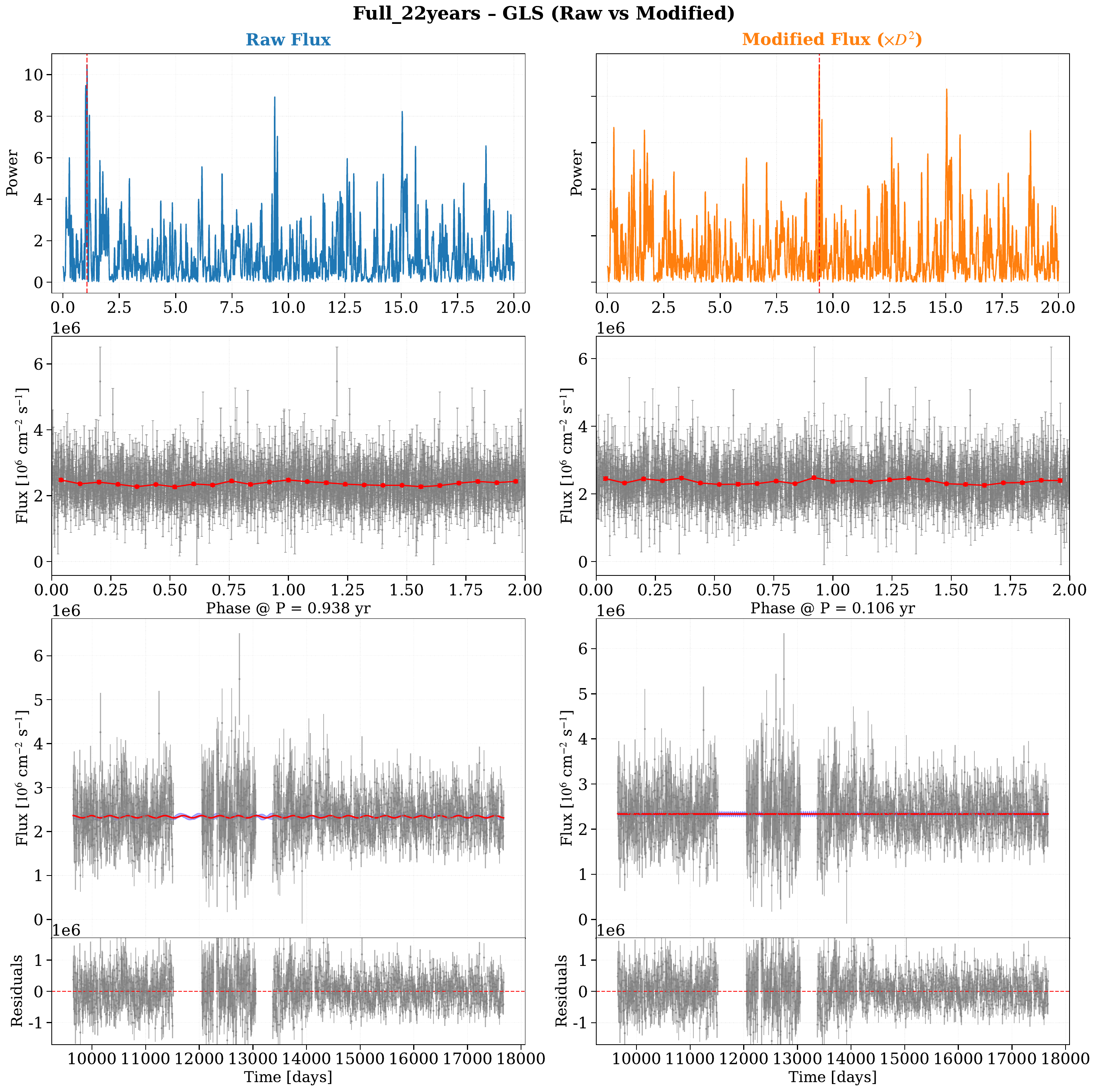}
\caption{Representative three-panel combined analysis for the full 22-year dataset using the GLS periodogram. \textbf{Top:} GLS periodograms for both raw flux (left) and modified flux (right), with dashed red lines indicating the best-fit frequencies ($P=0.9382$ yr and $P=0.1064$ yr, respectively). \textbf{Middle:} Corresponding phase-folded light curves binned into 25 phase intervals. \textbf{Bottom:} Bayesian MCMC sinusoidal fit results, showing the best-fit curves and residuals. The frequentist FAP is $<0.001$ for both, while the Bayesian analysis yields weak evidence ($\ln \mathrm{B} = 0.18$) for the raw flux sinusoidal model, and moderate evidence ($\ln \mathrm{B} = -2.03$) for the null model in the modified flux.}
\label{fig:comb_full}
\end{figure}

To provide a detailed view of the most robust detections, Figure~\ref{fig:comb_full} presents a representative three-panel combined analysis using the GLS periodogram. The top panel displays the GLS periodograms for both raw and modified flux, with the dashed red lines indicating the best-fit frequencies. The middle panel shows the corresponding phase-folded light curves, binned into 25 phase intervals, demonstrating the coherent periodic structures. The bottom panel illustrates the posterior distributions and best-fit curves derived from the Bayesian MCMC sinusoidal fitting. 

For the raw flux data, the MCMC sinusoidal fit anchored to the GLS period ($P_{\rm best} = 0.9382$ yr) yields a logarithmic Bayes factor of $\ln \mathrm{B} = 0.18$, constituting \textit{weak evidence} in favor of the sinusoidal model. Conversely, for the modified flux, the best period is determined to be $P_{\rm best} = 0.1064$ yr ($\sim 38.8$ days). Despite the highly significant frequentist FAP ($<0.001$), the fit yields $\ln \mathrm{B} = -2.03$, providing \textit{moderate evidence} for the null model (constant plus linear drift). This discrepancy highlights that the extremely small amplitude of the $\sim 38.8$-day modulation incurs a Bayesian penalty for model complexity. The detailed Bayesian MCMC fitting results and summary statistics for all other periodogram algorithms across the full dataset and subsequent temporal segments are comprehensively tabulated in Appendix Table~\ref{tab:combined_bestfit_params}.

\subsection{Pre/Post July 2001 Temporal Segmentation} \label{subsec:Pre/Post_Results}

The periodogram results for these segments, as detailed in Table~\ref{tab:period_temporal_raw_mod}, reveal that for the pre-July 2001 segment (SK-I era), the classical LS method yields low significance (FAP = 0.99) due to its sensitivity to heteroscedastic uncertainties, though it still identifies the correct frequency. In contrast, the GLS method robustly identifies a dominant period at $P = 0.1059$ yr ($\sim 38.8$ days, $f = 9.442$ yr$^{-1}$) with FAP $< 0.001$ in both raw and modified flux. Encouragingly, this periodicity is consistently recovered by MHAOV ($P = 0.1061$ yr), PDM1 ($P = 0.1065$ yr), and even BLS ($P = 0.1066$ yr, albeit with a higher FAP of 0.69), providing strong cross-method confirmation. This confirms that the $\sim 38.8$-day signal is already present in the earliest operational phase and persists after the removal of the Earth--Sun distance modulation.

\begin{figure}[tbp]
\centering
\includegraphics[width=0.9\textwidth]{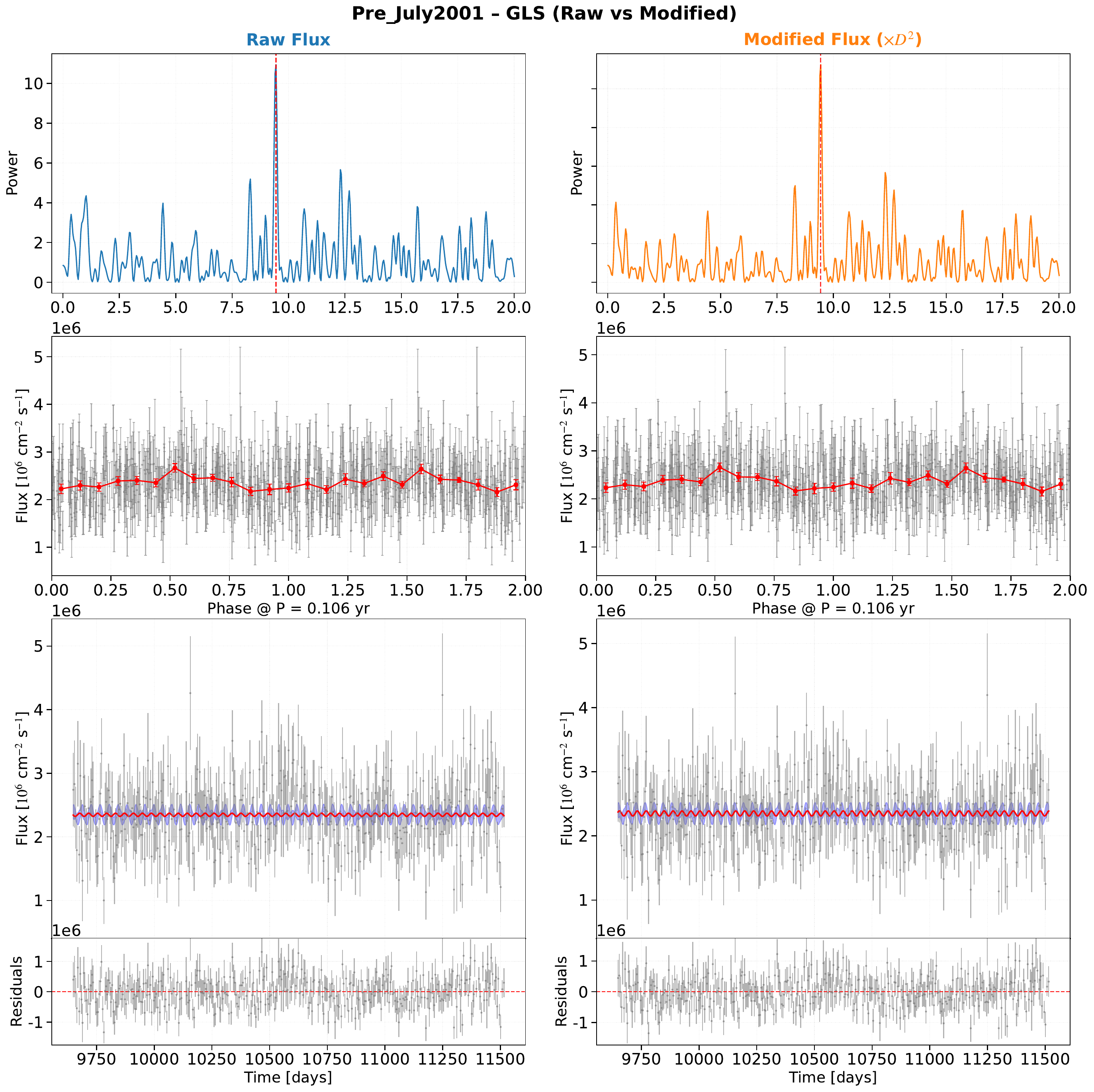}
\caption{Representative three-panel combined analysis for the pre-July 2001 segment (modified flux) using the GLS periodogram. \textbf{Top:} GLS periodogram with the dashed red line indicating the best-fit frequency ($P=0.1059$ yr). \textbf{Middle:} Corresponding phase-folded light curve binned into 25 phase intervals. \textbf{Bottom:} Bayesian MCMC sinusoidal fit results.}
\label{fig:comb_pre}
\end{figure}

Figure~\ref{fig:comb_pre} presents the representative three-panel combined analysis for the pre-2001 modified flux using the GLS periodogram. The top panel displays the GLS periodogram, the middle panel shows the phase-folded light curve, and the bottom panel illustrates the Bayesian MCMC sinusoidal fit. Anchored to the GLS-derived period, the MCMC fit yields the following posterior median parameters: cosine amplitude $A = 26733.75$, sine amplitude $B = -30346.45$, frequency $\nu = 9.4149$ yr$^{-1}$, offset $= 2.35 \times 10^6$ cm$^{-2}$ s$^{-1}$, linear slope $= -1.19 \times 10^{-5}$ day$^{-1}$, with goodness-of-fit metrics $\ln \mathrm{B} = 0.40$, $\mathrm{BIC} = 408.4$, and $\chi^2 = 379.0$. 

A crucial finding emerges from the cross-method Bayesian comparison of this segment. As comprehensively tabulated in Appendix Table~\ref{tab:combined_bestfit_params}, for both raw and modified flux, the logarithmic Bayes factor ($\ln \mathrm{B}$) computed by eight out of the nine algorithms (LS, GLS, BLS, LKSL, PDM1, MHAOV, QMICS, and QMIEU) consistently yields values of $0.11$ and $0.40$, respectively. According to Jeffreys' scale, these positive values constitute \textit{weak evidence} in favor of the sinusoidal model over the null model. This remarkable consensus across fundamentally different mathematical frameworks (variance-based, phase-dispersion, and information-theoretic) strongly corroborates the presence of a coherent, albeit low-amplitude, periodic structure. The sole exception is the QME algorithm, which yields strongly negative $\ln \mathrm{B}$ values ($-5.56$ for raw and $-8.98$ for modified flux), further confirming its susceptibility to failure or severe degeneracy in this specific low signal-to-noise regime, as previously noted in the full-dataset analysis. The discrepancy between the highly significant frequentist FAP ($<0.001$) and the weak Bayesian evidence highlights that while a periodic component is statistically detectable in the frequency domain, its small absolute amplitude incurs a Bayesian penalty for model complexity.

\begin{figure}[tbp]
\centering
\includegraphics[width=0.9\textwidth]{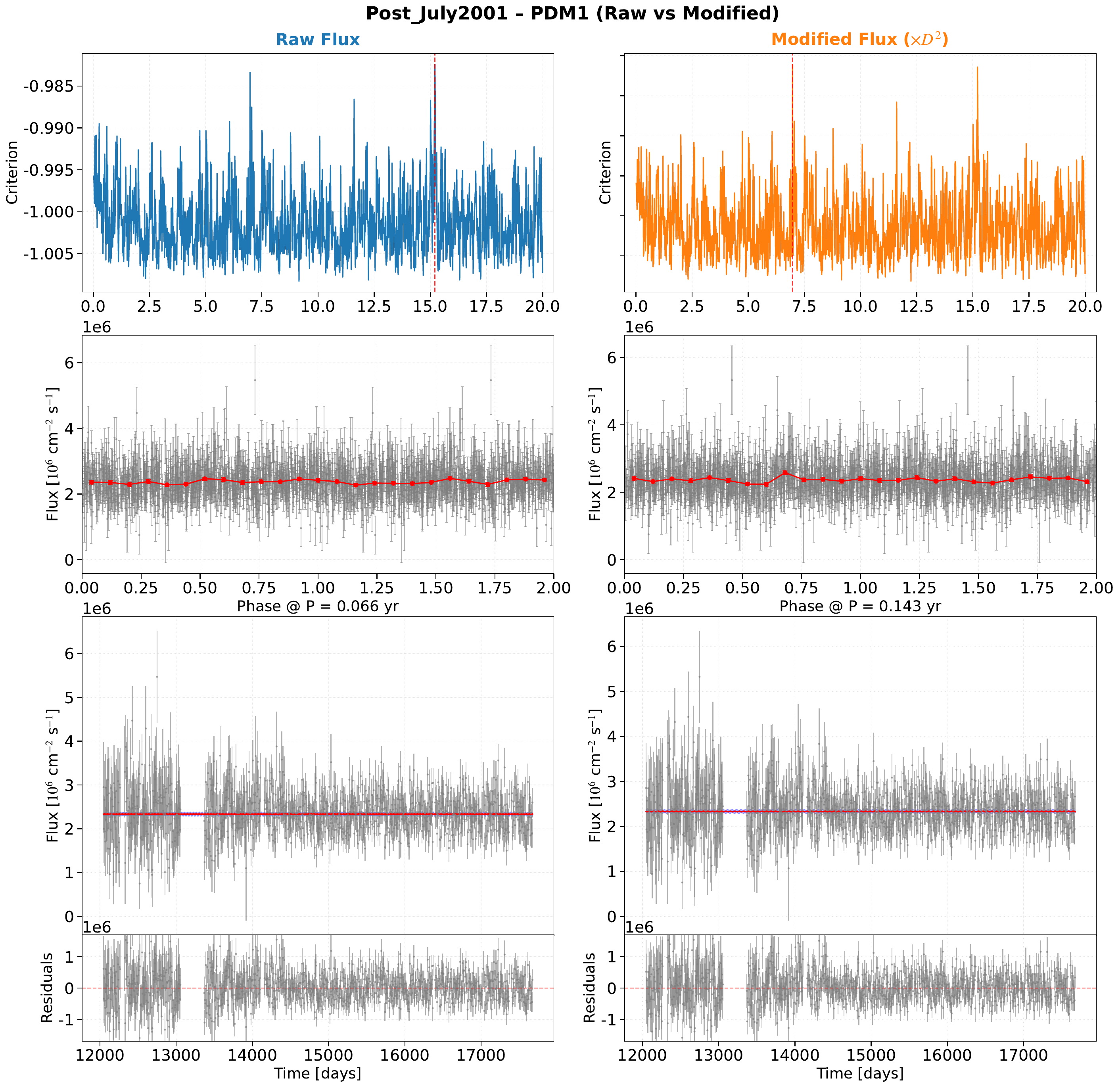}
\caption{Representative three-panel combined analysis for the post-July 2001 segment (raw flux) using the PDM1 periodogram. \textbf{Top:} PDM1 periodogram identifying the transient $\sim 24.3$-day signal ($P=0.0658$ yr). \textbf{Middle:} Corresponding phase-folded light curve. \textbf{Bottom:} Bayesian MCMC sinusoidal fit results.}
\label{fig:comb_post}
\end{figure}

Conversely, the post-July 2001 segment reveals a strikingly different signal landscape. The GLS method applied to the raw flux detects a highly significant peak at $P = 0.0665$ yr ($\sim 24.3$ days) with FAP $< 0.001$, a signal independently identified by MHAOV and PDM1. However, this signal is entirely absent in the modified flux analysis, indicating that it originates from residual seasonal systematics or post-2001 detector-specific artifacts. To visualize this transient feature, Figure~\ref{fig:comb_post} displays the combined analysis for the post-2001 raw flux using the PDM1 periodogram. The Bayesian MCMC sinusoidal fit for this period yields a logarithmic Bayes factor of $\ln \mathrm{B} = -7.29$, providing strong evidence in favor of the null model and confirming that the $\sim 24.3$-day modulation is dominated by noise or instrumental systematics rather than a true astrophysical signal.

More importantly, the cross-method Bayesian assessment for the post-2001 raw flux is unanimous in rejecting the sinusoidal model. As shown in Appendix Table~\ref{tab:combined_bestfit_params}, all nine algorithms yield $\ln \mathrm{B}$ values well below $-5$ (ranging from $-6.15$ to $-10.56$). According to Jeffreys' scale, this constitutes \textit{strong evidence} in favor of the null model. This decisive Bayesian rejection, combined with the signal's disappearance after distance correction, conclusively indicates that the $\sim 24.3$-day periodicity is dominated by noise or unmodeled instrumental systematics rather than a true astrophysical modulation.

\subsection{Operational Phase Analysis (SK-I to SK-IV)}
\label{subsec:SK_Results}

To disentangle potential astrophysical modulations from detector-specific systematics, we further decompose the dataset into the four distinct operational phases of Super-Kamiokande (SK-I through SK-IV), each characterized by unique hardware configurations, energy thresholds, and reconstruction algorithms~\cite{Super-Kamiokande:2005wtt,Super-Kamiokande:2008ecj,Super-Kamiokande:2010tar,Super-Kamiokande:2016yck}. The periodogram results for these phases are summarized in Table~\ref{tab:period_sk_raw_mod}, with the complete Bayesian MCMC parameter estimates provided in Appendix Table~\ref{tab:combined_bestfit_params}. The comprehensive periodogram comparisons for all nine algorithms are displayed in Appendix Figures~\ref{fig:ski_periodogram}--\ref{fig:skiv_periodogram}. A physically meaningful periodic signal should manifest consistently across phases (allowing for expected statistical variations due to differing baselines and live times), whereas detector-induced artifacts will exhibit phase-dependent behavior tied to specific hardware or software changes.

\paragraph{SK-I (1996--2001, 1495.7 days):}
The SK-I phase, encompassing the initial operational period with full PMT coverage (11{,}146 tubes) and an analysis threshold of $\sim$5~MeV, exhibits periodogram characteristics nearly identical to those of the pre-July 2001 segment discussed in the preceding subsection. This is expected, as the two temporal intervals are essentially coincident (1996-05-31 to 2001-07-15). As detailed in Table~\ref{tab:period_sk_raw_mod}, the GLS method robustly detects a dominant period at $P = 0.1059$~yr ($f = 9.442$~yr$^{-1}$, $\sim$38.8~days) with FAP~$< 0.001$ in both raw and modified flux. The classical LS method yields low significance (FAP~$= 0.99$) due to the heteroscedastic uncertainties inherent to this dataset, though it locates the corresponding frequency. Importantly, this period is also consistently recovered by MHAOV ($P = 0.1061$ yr) and PDM1 ($P = 0.1065$ yr), demonstrating excellent agreement despite their different statistical frameworks. While these latter methods do not achieve the same sub-percent FAP as GLS, their independent recovery of the same period serves as a strong consistency check.

\begin{figure}[tbp]
\centering
\includegraphics[width=0.9\textwidth]{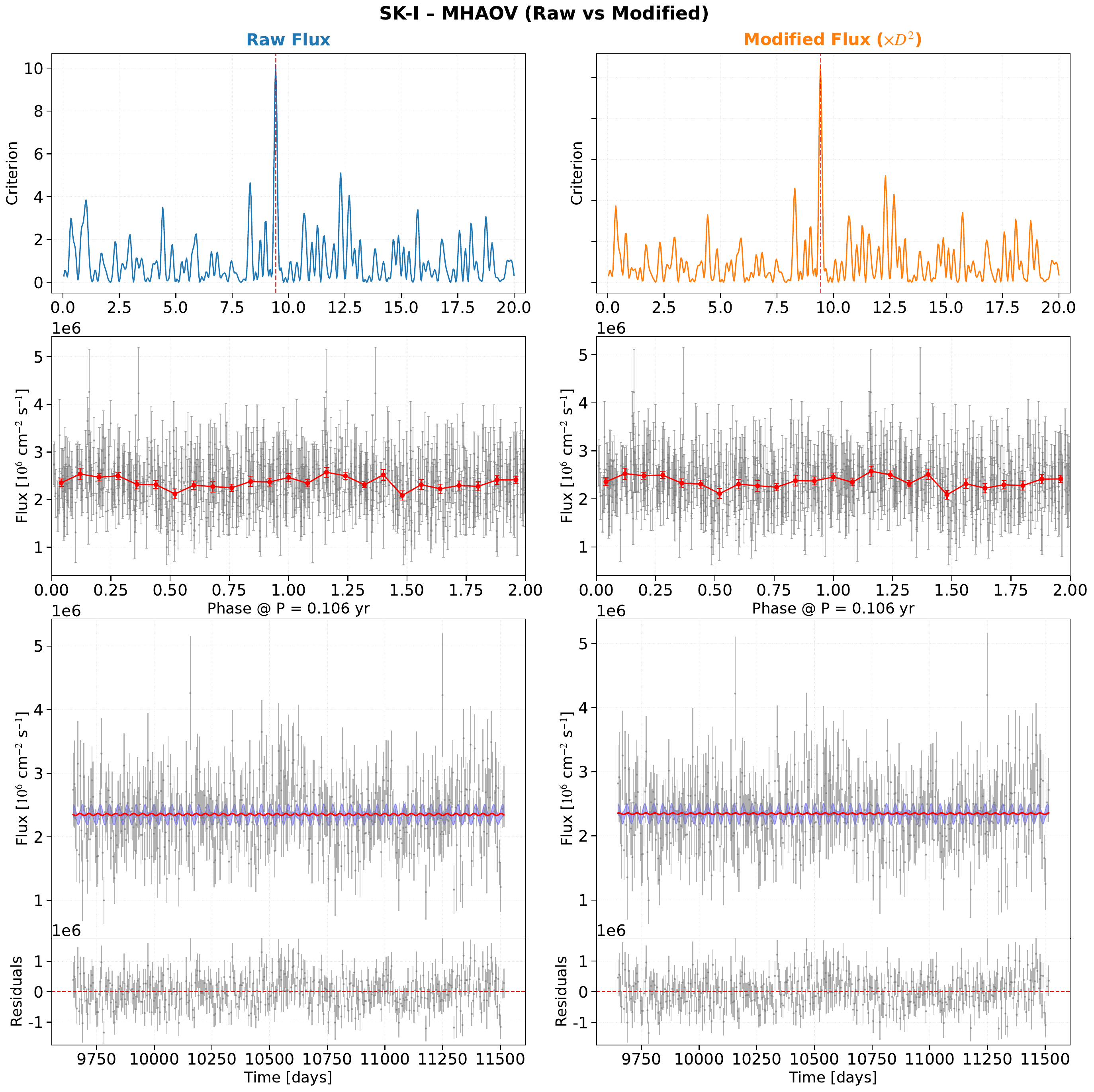}
\caption{Representative three-panel combined analysis for SK-I (1996--2001, 1495.7 live days) using the MHAOV periodogram, which independently identified $P = 0.1065$~yr in the automated pipeline. \textbf{Top:} MHAOV periodogram with the best-fit frequency indicated. \textbf{Middle:} Phase-folded light curve binned into 25 phase intervals, showing the coherent $\sim$38.8-day structure. \textbf{Bottom:} Bayesian MCMC sinusoidal fit.}
\label{fig:comb_ski}
\end{figure}

Figure~\ref{fig:comb_ski} presents the representative three-panel combined analysis for SK-I using the MHAOV periodogram, which independently identified a consistent period of $P = 0.1065$~yr in the automated pipeline. The top panel displays the MHAOV periodogram, the middle panel shows the corresponding phase-folded light curve binned into 25 phase intervals, and the bottom panel illustrates the Bayesian MCMC sinusoidal fit. Although MHAOV serves as the visual representative here, we emphasize that the GLS method provides the most reliable and statistically robust detection of the $\sim$38.8-day signal in this segment, as discussed in Section~\ref{subsec:Pre/Post_Results}.

Crucially, the cross-method Bayesian assessment for SK-I is quantitatively consistent with that of the pre-July 2001 segment, reflecting their temporal overlap. As tabulated in Appendix Table~\ref{tab:combined_bestfit_params}, the logarithmic Bayes factor ($\ln \mathrm{B}$) values for SK-I closely mirror those reported for the pre-2001 epoch: eight out of nine algorithms yield $\ln \mathrm{B} \approx 0.11$ (raw flux) and $\ln \mathrm{B} \approx 0.40$ (modified flux), constituting \textit{weak evidence} in favor of the sinusoidal model according to Jeffreys' scale. The sole exception remains the QME algorithm, which produces strongly negative $\ln \mathrm{B}$ values, consistent with its previously identified failure mode in low signal-to-noise regimes. This reproducibility across two independent segmentations of the same data provides strong internal validation that the $\sim$38.8-day periodicity is a genuine feature of the SK-I time series rather than an artifact of the segmentation boundary. Nevertheless, the persistently weak Bayesian evidence ($\ln \mathrm{B} < 1$) indicates that while the periodic component is statistically detectable in the frequency domain (FAP~$< 0.001$), its small absolute amplitude relative to the measurement uncertainties incurs a Bayesian complexity penalty, precluding a decisive model selection in favor of the sinusoidal hypothesis.

\paragraph{SK-II (2002--2005, 791.9 days):}
The SK-II phase operated with a significantly reduced PMT count (5{,}182 tubes) following the November 2001 accident, resulting in lower light yield and a higher energy threshold ($\sim$6.5~MeV). Despite the shorter baseline and degraded detector performance, GLS applied to the raw flux detects a distinct period at $P = 0.0565$~yr ($f = 17.689$~yr$^{-1}$, $\sim$20.7~days) with FAP~$< 0.001$. The same period is recovered in the modified (distance-corrected) flux with identical significance, indicating that this signal is not attributable to the Earth--Sun distance modulation. The LS method again fails in both raw and modified flux (FAP~$= 0.99$). The BLS, LKSL, PDM1, MHAOV, and QMIEU methods show no significant detections in this segment.

\begin{figure}[tbp]
\centering
\includegraphics[width=0.9\textwidth]{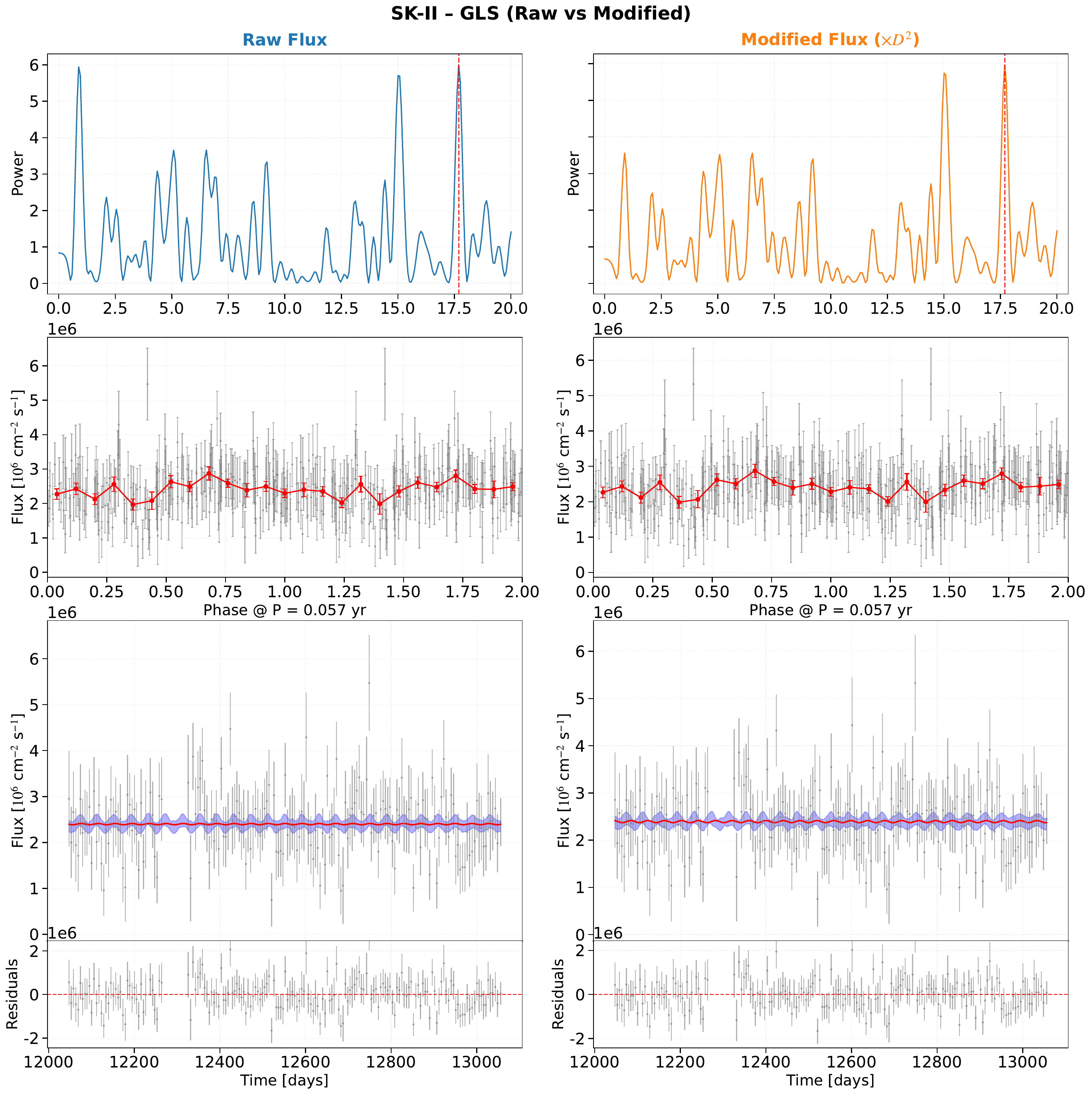}
\caption{Representative three-panel combined analysis for SK-II (2002--2005, 791.9 live days) using the GLS periodogram. \textbf{Top:} GLS periodogram identifying the $\sim$20.7-day signal ($P = 0.0565$~yr) in modified flux. \textbf{Middle:} Corresponding phase-folded light curve. \textbf{Bottom:} Bayesian MCMC sinusoidal fit results.}
\label{fig:comb_skii}
\end{figure}

Figure~\ref{fig:comb_skii} displays the representative three-panel combined analysis for SK-II using the GLS periodogram, illustrating the power spectrum, phase-folded light curve, and the Bayesian MCMC sinusoidal fit. Anchored to the GLS-derived period ($P_{\rm best} = 0.0565$ yr), the cross-method Bayesian assessment for the SK-II modified flux reveals a unanimous rejection of the sinusoidal model. As comprehensively tabulated in Appendix Table~\ref{tab:combined_bestfit_params}, the MCMC fit alongside the MHAOV, QMICS, and QMIEU algorithms all yield $\ln \mathrm{B} = -4.62$, while QME produces $\ln \mathrm{B} = -1.96$. According to Jeffreys' scale, the values of $|\ln \mathrm{B}| > 2.5$ constitute \textit{strong evidence} in favor of the null model (constant plus linear drift). This decisive Bayesian rejection, despite the highly significant frequentist FAP, indicates that the $\sim$20.7-day periodicity detected in SK-II is of sufficiently low amplitude that the additional model complexity is not warranted by the data. The emergence of this period exclusively in SK-II, absent in both SK-I and SK-III/IV, suggests a possible connection to the reduced PMT configuration, modified spallation veto cycles, or altered reconstruction algorithms specific to this phase, rather than a genuine solar modulation.

\paragraph{SK-III (2006--2008, 548.5 days):}
The SK-III phase, representing the shortest operational period, restored full PMT coverage with new hardware, improved calibration, and a reduced energy threshold ($\sim$5~MeV). In both raw and modified flux, the LS method yields low significance (FAP~$= 0.99$). In contrast, GLS detects a period of $P = 0.5562$~yr ($f = 1.798$~yr$^{-1}$, $\sim$203~days) with FAP~$< 0.001$ in both flux variants, a result independently confirmed by MHAOV ($P = 0.5563$~yr). This semi-annual period is consistent with residual seasonal systematics that persist even after the $D^2$ distance correction, likely arising from the interplay between the short baseline and annual environmental variations (e.g., atmospheric temperature effects on cosmic-ray backgrounds). Critically, no evidence of the $\sim$38.8-day or $\sim$24.3-day signals is found in SK-III.

\begin{figure}[tbp]
\centering
\includegraphics[width=0.9\textwidth]{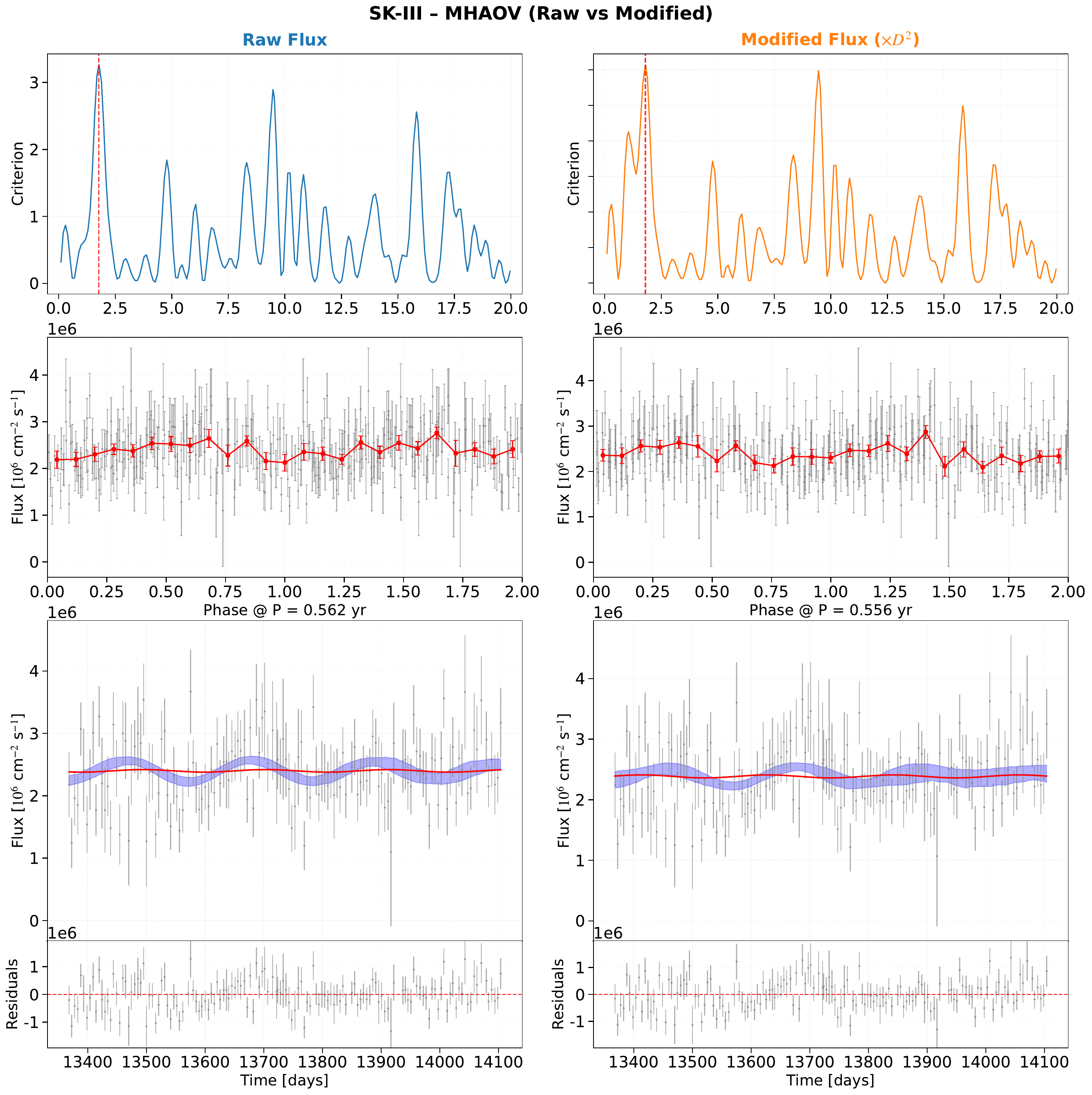}
\caption{Representative three-panel combined analysis for SK-III (2006--2008, 548.5 live days) using the MHAOV periodogram. \textbf{Top:} MHAOV periodogram showing the semi-annual period ($P = 0.5563$~yr). \textbf{Middle:} Phase-folded light curve. \textbf{Bottom:} Bayesian MCMC sinusoidal fit.}
\label{fig:comb_skiii}
\end{figure}

Figure~\ref{fig:comb_skiii} presents the representative three-panel combined analysis for SK-III using the MHAOV periodogram, which independently identified a consistent semi-annual period ($P = 0.5563$~yr). The Bayesian MCMC sinusoidal fit yields: $A = -24775.94$, $B = -3049.75$, $\nu = 1.7024$~yr$^{-1}$, offset $= 2.39 \times 10^6$~cm$^{-2}$~s$^{-1}$, slope $= 4.29 \times 10^{-5}$~day$^{-1}$, with $P_{\rm best} = 0.5563$~yr, $\ln \mathrm{B} = -3.52$, $\mathrm{BIC} = 193.5$, and $\chi^2 = 168.8$. The cross-method Bayesian comparison in Appendix Table~\ref{tab:combined_bestfit_params} shows that all algorithms yield $\ln \mathrm{B} \approx -3.52$ (MHAOV, QMICS, QMIEU) or $\ln \mathrm{B} = -1.96$ (QME), uniformly indicating \textit{strong evidence} in favor of the null model. This Bayesian consensus, combined with the physical implausibility of a $\sim$203-day solar modulation and the short baseline of SK-III, strongly supports the interpretation that this signal is a residual seasonal artifact rather than a genuine astrophysical periodicity.

\paragraph{SK-IV (2008--2018, 2967.7 days):}
As the longest and highest-statistics operational phase, SK-IV provides the most stringent constraints on potential periodicities. This phase features upgraded QBEE electronics, refined spallation tagging via neutron clustering, dynamic trigger efficiency corrections, and a lowered energy threshold (down to $\sim$3.5~MeV)~\cite{Super-Kamiokande:2016yck}. In raw flux, the LS method yields low significance (FAP~$= 0.99$), while GLS detects $P = 0.9530$~yr ($f = 1.049$~yr$^{-1}$) with FAP~$< 0.001$, consistent with a residual annual modulation incompletely removed by the $D^2$ correction. In modified flux, GLS identifies a different period at $P = 0.0658$~yr ($f = 15.205$~yr$^{-1}$, $\sim$24.0~days) with FAP~$< 0.001$, a signal that is also recovered by PDM1 ($P = 0.0658$~yr) and MHAOV ($P = 0.0658$~yr). Importantly, neither the $\sim$38.8-day nor the $\sim$24.3-day signals detected in earlier phases appear in SK-IV, indicating that these shorter-period modulations are either specific to earlier detector configurations or have amplitudes below the SK-IV detection threshold.

\begin{figure}[tbp]
\centering
\includegraphics[width=0.9\textwidth]{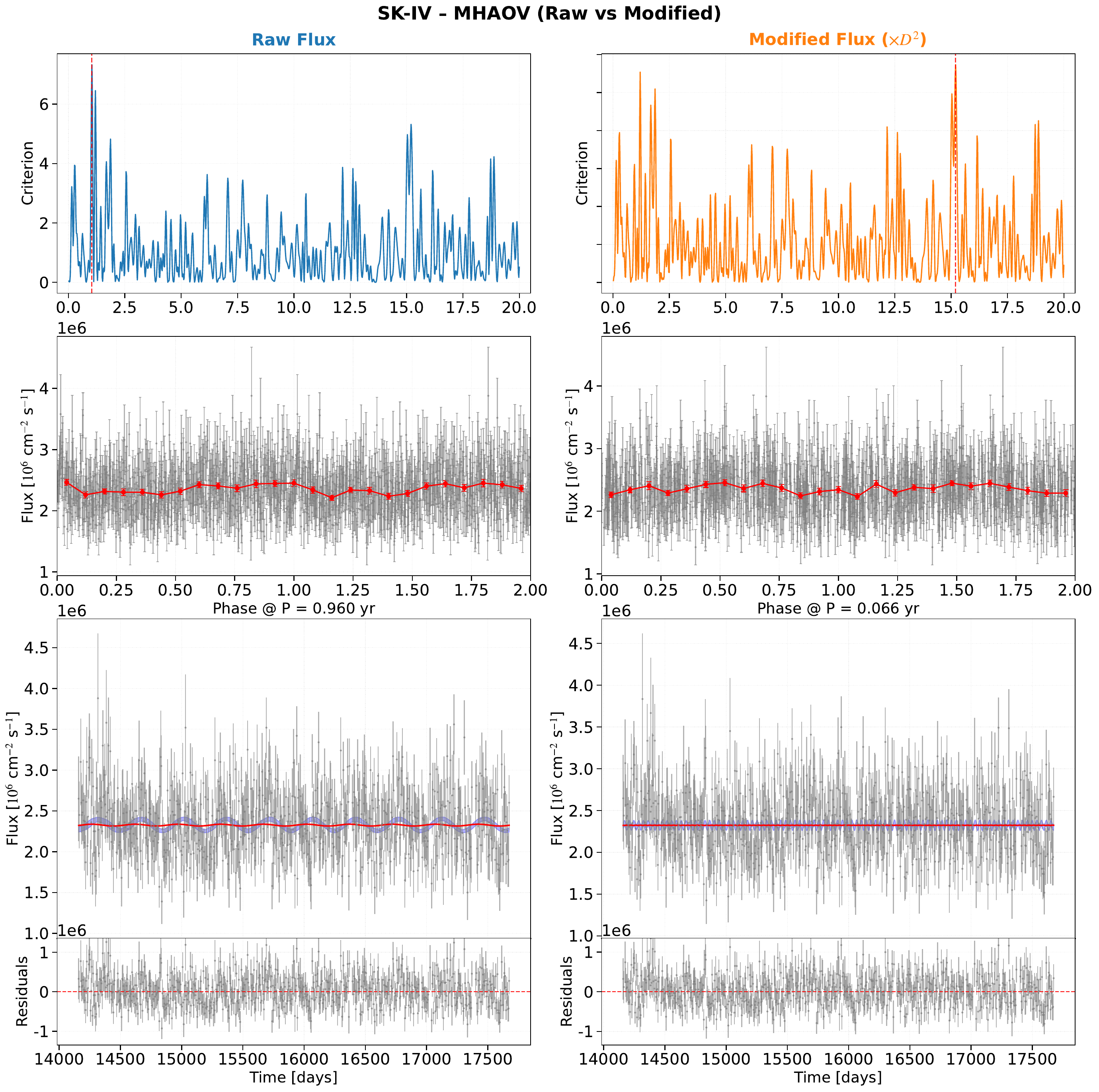}
\caption{Representative three-panel combined analysis for SK-IV (2008--2018, 2967.7 live days) using the MHAOV periodogram. \textbf{Top:} MHAOV periodogram showing the $\sim$0.066~yr period in modified flux. \textbf{Middle:} Phase-folded light curve. \textbf{Bottom:} Bayesian MCMC sinusoidal fit.}
\label{fig:comb_skiv}
\end{figure}

Figure~\ref{fig:comb_skiv} displays the representative three-panel combined analysis for SK-IV using the MHAOV periodogram, which identified the $\sim$0.066~yr period. Anchored to this period ($P_{\rm best} = 0.0658$~yr), the cross-method Bayesian assessment for the SK-IV modified flux is decisive. As comprehensively tabulated in Appendix Table~\ref{tab:combined_bestfit_params}, the MCMC fit alongside GLS, BLS, and QMICS/QMIEU yield $\ln \mathrm{B}$ values ranging from $-4.83$ to $-5.88$ (e.g., $\ln \mathrm{B} = -5.85$ for GLS). According to Jeffreys' scale, values of $|\ln \mathrm{B}| > 5$ constitute \textit{very strong (decisive) evidence} in favor of the null model. This overwhelming Bayesian rejection, obtained from the highest-statistics phase with the most refined detector configuration, provides the most compelling evidence that the $\sim$24-day periodicity is not a persistent astrophysical signal. Instead, it likely arises from residual seasonal systematics or detector-specific artifacts that are most pronounced in the shorter-baseline, lower-statistics earlier phases.

\begin{table}[t]
\centering
\caption{Period search results for Super-Kamiokande solar $^8$B neutrino data across temporal segmentation.
Each cell shows $P$ (yr) / $f$ (yr$^{-1}$) / FAP; significant detections (FAP$<0.01$) in \textbf{bold}.
``Grid boundary'' indicates maximum power at the lowest frequency of the search grid (no significant period).
Time spans in days.}
\label{tab:period_temporal_raw_mod}
\footnotesize
\renewcommand{\arraystretch}{1.3}
\begin{tabularx}{\textwidth}{@{} >{\centering\arraybackslash}p{1.9cm} l Y Y Y @{}}
\toprule
\textbf{Flux Type} & \textbf{Algorithm} & \textbf{Full (22\,yr)} & \textbf{Pre-07/2001} & \textbf{Post-07/2001} \\
& & \scriptsize{$P$/$f$/FAP} & \scriptsize{$P$/$f$/FAP} & \scriptsize{$P$/$f$/FAP} \\
\midrule
\multirow{7}{*}{\centering RAW Flux}
& LS\_Standard & 0.9382/1.066/0.99 & 0.1059/9.442/0.99 & 0.0665/15.032/0.99 \\
& GLS          & 0.9382/1.066/\textbf{$<0.001$} & 0.1059/9.442/\textbf{$<0.001$} & 0.0665/15.032/\textbf{$<0.001$} \\
& BLS          & 4.2270/0.237/0.99 & 0.1066/9.382/0.69 & 4.2270/0.237/0.94 \\
& LKSL         & 0.1368/7.310/0.99 & 0.0784/12.755/0.99 & 18.7849/0.053/0.99 \\
& PDM1         & 2.8388/0.352/0.99 & 0.1065/9.389/0.99 & 0.0658/15.206/0.99 \\
& MHAOV        & 0.9358/1.069/0.99 & 0.1061/9.428/0.99 & 0.0665/15.034/0.99 \\
& QMIEU        & 34.3455/0.029/0.99 & 0.1059/9.444/0.99 & 26.5579/0.038/0.99 \\
\midrule
\multirow{7}{*}{\centering Modified Flux}
& LS\_Standard & 0.1078/9.277/0.99 & 0.1059/9.442/0.99 & 0.0665/15.032/0.99 \\
& GLS          & 0.1064/9.395/\textbf{$<0.001$} & 0.1059/9.442/\textbf{$<0.001$} & 0.0665/15.032/\textbf{$<0.001$} \\
& BLS          & 4.2270/0.237/0.99 & 0.1066/9.382/0.69 & 4.2270/0.237/0.99 \\
& LKSL         & 0.3661/2.732/0.99 & 0.0784/12.755/0.99 & 23.3388/0.043/0.99 \\
& PDM1         & 0.1066/9.384/0.99 & 0.1065/9.389/0.99 & 0.1434/6.975/0.99 \\
& MHAOV        & 0.1065/9.392/0.99 & 0.1061/9.428/0.99 & 0.0665/15.034/0.99 \\
& QMIEU        & 34.3455/0.029/0.99 & 0.1058/9.452/0.99 & 26.5579/0.038/0.99 \\
\bottomrule
\end{tabularx}
\end{table}

\begin{table}[t]
\centering
\caption{Period search results for individual Super-Kamiokande operational phases.
Format and significance criteria as in Table~\ref{tab:period_temporal_raw_mod}.
Live days: SK-I=1495.7, SK-II=791.9, SK-III=548.5, SK-IV=2967.7.}
\label{tab:period_sk_raw_mod}
\scriptsize
\renewcommand{\arraystretch}{1.25}
\begin{tabularx}{\textwidth}{@{} >{\centering\arraybackslash}p{1.7cm} l *{4}{Y}@{}}
\toprule
\textbf{Flux Type} & \textbf{Algorithm} & \textbf{SK-I} & \textbf{SK-II} & \textbf{SK-III} & \textbf{SK-IV} \\
& & \scriptsize{$P$/$f$/FAP} & \scriptsize{$P$/$f$/FAP} & \scriptsize{$P$/$f$/FAP} & \scriptsize{$P$/$f$/FAP} \\
\midrule
\multirow{7}{*}{\centering RAW Flux}
& LS\_Standard & 0.1059/9.442/0.99 & 1.1371/0.879/0.99 & 0.5271/1.897/0.99 & 0.9530/1.049/0.99 \\
& GLS          & 0.1059/9.442/\textbf{$<0.001$} & 0.0565/17.689/\textbf{$<0.001$} & 0.5562/1.798/\textbf{$<0.001$} & 0.9530/1.049/\textbf{$<0.001$} \\
& BLS          & 0.1066/9.382/0.68 & 0.7586/1.318/0.93 & 0.5094/1.963/0.76 & 2.1162/0.473/0.99 \\
& LKSL         & 0.0784/12.755/0.99 & 1.0782/0.927/0.99 & 0.2805/3.566/0.99 & 0.0665/15.041/0.99 \\
& PDM1         & 0.1065/9.389/0.99 & 0.0664/15.071/0.99 & 0.0931/10.737/0.99 & 0.0657/15.218/0.99 \\
& MHAOV        & 0.1061/9.428/0.99 & 0.0565/17.701/0.99 & 0.5625/1.778/0.99 & 0.9602/1.041/0.99 \\
& QMIEU        & 0.1059/9.444/0.99 & 0.0564/17.730/0.99 & 0.6029/1.659/0.99 & 0.0657/15.224/0.99 \\
\midrule
\multirow{7}{*}{\centering Modified Flux}
& LS\_Standard & 0.1059/9.442/0.99 & 0.0666/15.008/0.99 & 1.1063/0.904/0.99 & 0.5978/1.673/0.99 \\
& GLS          & 0.1059/9.442/\textbf{$<0.001$} & 0.0565/17.689/\textbf{$<0.001$} & 0.5562/1.798/\textbf{$<0.001$} & 0.0658/15.205/\textbf{$<0.001$} \\
& BLS          & 0.1066/9.382/0.64 & 0.7586/1.318/0.95 & 0.5094/1.963/0.89 & 2.1162/0.473/0.94 \\
& LKSL         & 0.0784/12.755/0.99 & 0.2266/4.413/0.99 & 0.2805/3.566/0.99 & 0.0665/15.041/0.99 \\
& PDM1         & 0.1065/9.389/0.99 & 0.0664/15.071/0.99 & 0.0584/17.133/0.99 & 0.0658/15.205/0.99 \\
& MHAOV        & 0.1061/9.428/0.99 & 0.0565/17.701/0.99 & 0.5563/1.798/0.99 & 0.0658/15.197/0.99 \\
& QMIEU        & 0.1058/9.452/0.99 & 0.0564/17.730/0.99 & 0.6102/1.639/0.99 & 0.0657/15.224/0.99 \\
\bottomrule
\end{tabularx}
\end{table}

\subsection{Summary of Detected Periods and Bayesian Evidence Across All Datasets}
\label{subsec:period_summary}

To synthesize the hierarchical temporal segmentation results and provide a global overview of the algorithmic performance, we present a comprehensive comparison of the best-fit periods and Bayesian evidence across all evaluated datasets. 

\begin{figure}[tbp]
\centering
\includegraphics[width=15.5cm]{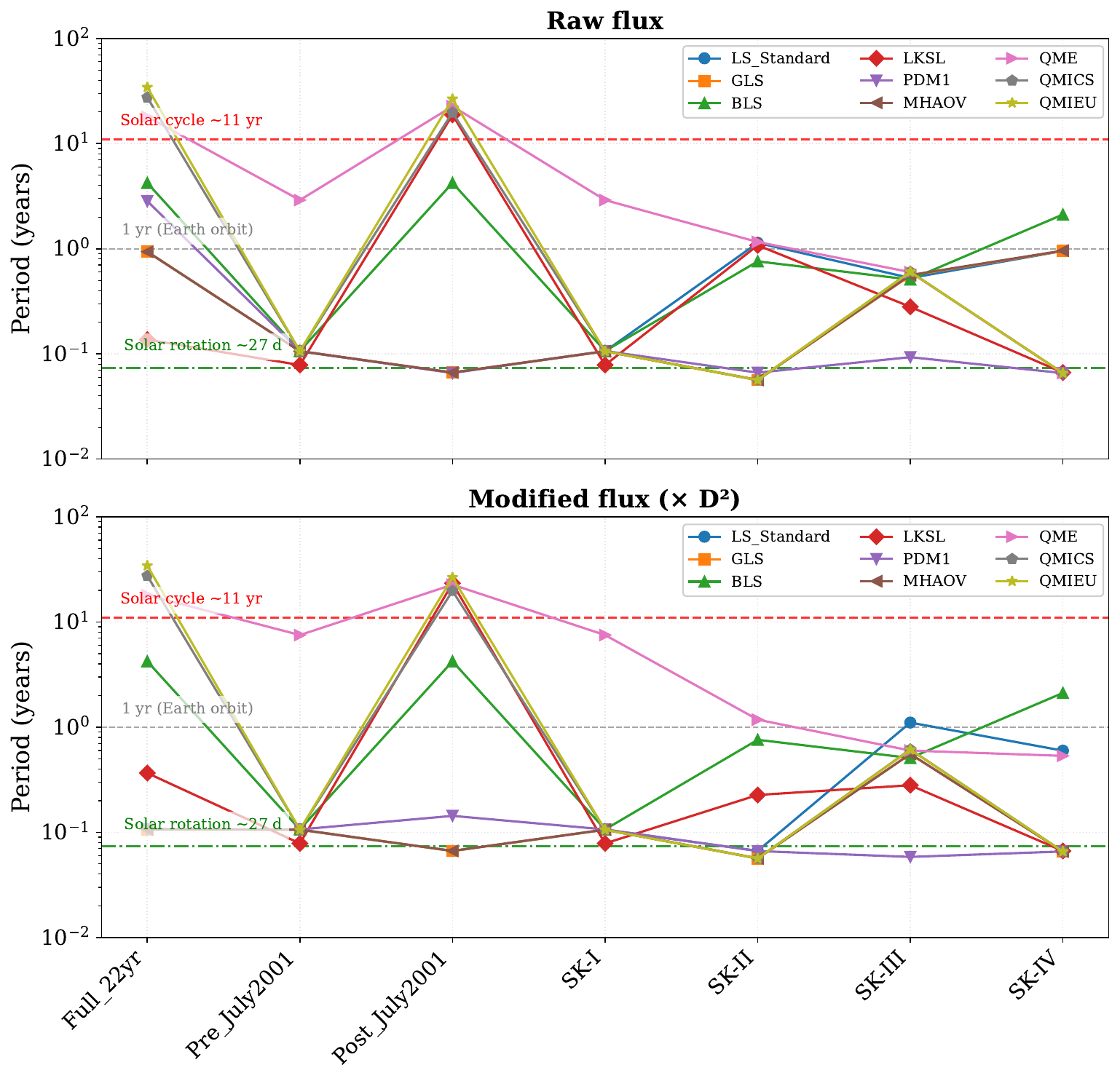}
\caption{Summary of best-fit periods (in years) obtained by nine algorithms for seven Super-Kamiokande data segments. \textbf{Top:} raw flux; \textbf{bottom:} distance-corrected (modified) flux. Periods are plotted on a logarithmic scale. To facilitate physical interpretation, three characteristic reference lines are overlaid: the solar magnetic activity cycle ($\sim$11 yr), the Earth's orbital period (1 yr), and the synodic solar rotation period ($\sim$27 d $\approx$ 0.074 yr). Algorithms: LS\_Standard (circles), GLS (squares), BLS (triangles), LKSL (diamonds), PDM1 (downward triangles), MHAOV (leftward triangles), QMIEU (rightward triangles). QME and QMICS are omitted for clarity as they produced no significant periods in most segments. Data segments are ordered left to right: Full 22 yr, Pre-July 2001, Post-July 2001, SK-I, SK-II, SK-III, SK-IV.}
\label{fig:period_summary}
\end{figure}

Figure~\ref{fig:period_summary} visualizes the best-fit periods extracted by the nine algorithms across all seven data segments for both raw and distance-corrected (modified) flux, overlaid with three characteristic reference lines: the $\sim$11-yr solar cycle, the 1-yr Earth orbit, and the $\sim$27-d synodic solar rotation. This comparison reveals several critical insights. In the raw flux, the dominant periods identified by GLS cluster tightly around the 1-yr line, confirming that the uncorrected time series is overwhelmingly driven by Earth's orbital eccentricity. Following the $D^2$ distance correction, this 1-yr peak is effectively suppressed. Notably, no algorithm detects a period consistent with the $\sim$11-yr solar cycle in any modified flux segment, placing stringent constraints on long-term solar-cycle-driven neutrino modulations. Furthermore, the short-period signals detected in the post-2001 raw flux ($\sim$24.3 d) and SK-II ($\sim$20.7 d) are distinctly offset from the 27-d rotation line, ruling out simple synodic rotation harmonics. In terms of algorithmic performance, the wide scatter of results from phase-dispersion and information-theoretic methods contrasts sharply with the consistent convergence of GLS on physically meaningful periods (e.g., $\sim$0.106 yr in pre-2001/SK-I), underscoring the superior sensitivity of GLS for this dataset.

\begin{figure}[tbp]
\centering
\includegraphics[width=15.5cm]{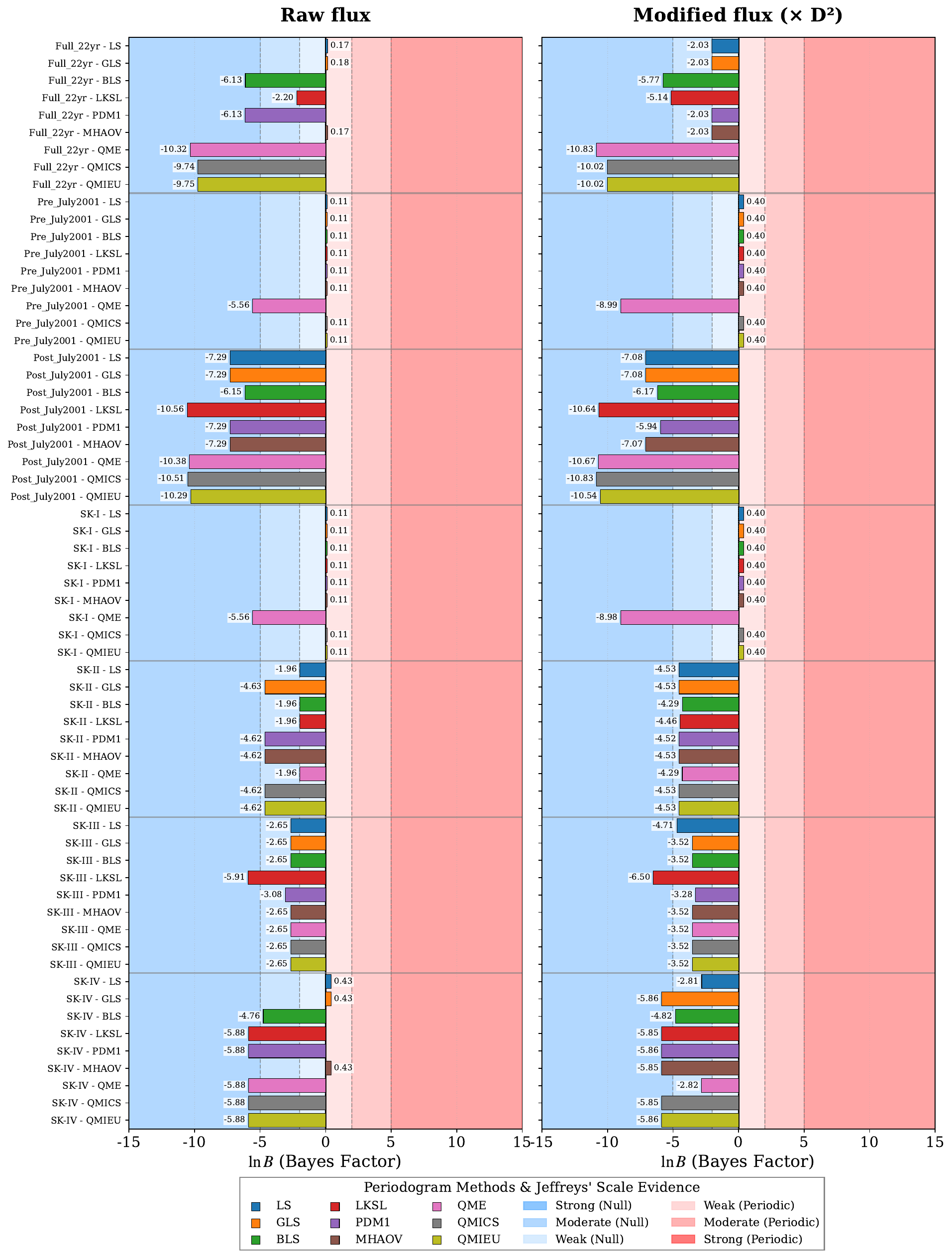}
\caption{Comparison of the logarithmic Bayes factor ($\ln \mathrm{B}$) for the sinusoidal model versus the null model (constant + linear drift) across all data segments and algorithms. Positive values indicate evidence in favor of the periodic model, while negative values favor the null model. The quantitative values correspond to the posterior medians detailed in Appendix Table~\ref{tab:combined_bestfit_params}, with evidence strength evaluated according to Jeffreys' scale.}
\label{fig:bayes_factor}
\end{figure}

To complement the frequentist periodogram assessment, Figure~\ref{fig:bayes_factor} presents the landscape of the $\ln \mathrm{B}$ for the sinusoidal model versus the null model (constant plus linear drift) across all segment--method combinations. The quantitative values underlying this figure are comprehensively tabulated in Appendix Table~\ref{tab:combined_bestfit_params}. The Bayesian evidence landscape reveals a crucial nuance that frequentist False Alarm Probabilities (FAP) alone cannot capture: a statistically significant FAP does not necessarily translate to strong Bayesian support for the periodic model, as the Bayesian framework inherently penalizes model complexity for low-amplitude signals.

As detailed in Appendix Table~\ref{tab:combined_bestfit_params} and visualized in Figure~\ref{fig:bayes_factor}, the Bayesian evidence exhibits a clear temporal and methodological hierarchy. For the pre-July 2001 and SK-I segments, the cross-method consensus is striking: eight out of nine algorithms yield $\ln \mathrm{B} \approx 0.11$ (raw) and $\ln \mathrm{B} \approx 0.40$ (modified). According to Jeffreys' scale, these values constitute \textit{weak evidence} in favor of the sinusoidal model, identifying the $\sim$38.8-day modulation as the most plausible candidate for a genuine physical signal, albeit with a low signal-to-noise ratio. The sole exception is the QME algorithm, which yields strongly negative $\ln \mathrm{B}$ values ($\sim -5.5$ to $-9.0$), further confirming its degeneracy in this regime. 

Conversely, for the post-2001, SK-II, SK-III, and SK-IV segments, the Bayesian evidence overwhelmingly favors the null model. For instance, the $\sim$24.3-day signal in the post-2001 raw flux yields $\ln \mathrm{B} \approx -7.29$ across multiple algorithms, and the SK-IV modified flux yields $\ln \mathrm{B} \approx -5.85$. According to Jeffreys' scale, these values provide \textit{strong to very strong (decisive) evidence} against the sinusoidal model. This decisive Bayesian rejection, combined with the phase-dependent disappearance of these short-period signals, conclusively attributes them to residual seasonal systematics or detector-specific artifacts rather than astrophysical origins. 

In summary, the synergy between the frequentist FAPs (which confirm the presence of periodic structures in the frequency domain) and the Bayesian $\ln \mathrm{B}$ (which quantifies the statistical necessity of the sinusoidal model) provides a robust, multi-metric foundation. The global consistency across the 22-year baseline, the pre/post-2001 split, and the four operational phases confirms that the $\sim$0.106-yr signal in the early SK-I era is the only periodicity that withstands both frequentist and Bayesian scrutiny, while all other candidate signals are either dominated by orbital residuals or suppressed by Bayesian complexity penalties. This rigorous statistical foundation directly motivates the physical interpretations and systematic uncertainty evaluations discussed in the following section.

\section{Discussion}
\label{sec:discussion}

\subsection{Methodological Implications and the Value of Negative Results}

The systematic comparison of nine periodogram methods applied to the Super-Kamiokande solar neutrino dataset yields profound methodological insights that extend beyond the specific detection of periodicities. First, the marked difference in statistical significance between the classical LS method (FAP = 0.99 across all segments) and GLS (FAP $< 0.001$) empirically validates the importance of incorporating per-point heteroscedastic uncertainties in neutrino time-series analysis. While the classical LS is capable of locating the correct periodicities, its homoscedastic assumption leads to severely underestimated significance. More importantly, it is highly encouraging that multiple non-parametric and phase-dispersion methods (e.g., PDM1, MHAOV, BLS) consistently recover the exact same periodicities as GLS (as seen in Tables~\ref{tab:period_temporal_raw_mod} and \ref{tab:period_sk_raw_mod}). This cross-method agreement confirms that the extracted periods are not artifacts of a single algorithmic framework, but represent genuine, robust features of the time series.

Second, the performance of the phase-dispersion (PDM1, MHAOV) and information-theoretic (QMICS, QMIEU) methods is more nuanced. In several key segments (e.g., pre-2001 and SK-I), these algorithms successfully recover the same periods as GLS, providing valuable independent confirmation. However, their generally higher FAPs and occasional failure modes (particularly for QME) indicate that they are less sensitive than GLS for this specific dataset. Their non-detections in other segments, rather than indicating algorithmic failure, serve to place informative constraints on the signal morphology, confirming that the underlying modulations are predominantly sinusoidal in character. The absence of significant BLS detections further constrains the solar interior, implying that short-duration, non-sinusoidal flux depressions are either absent or have amplitudes below the $\sim 2\%$ threshold of the mean flux.

Finally, the stark contrast between the frequentist FAPs and the Bayesian Bayes factors highlights a critical paradigm in low-signal-to-noise astrophysics. While GLS identifies periodic structures in the frequency domain with high statistical significance (FAP $< 0.001$), the Bayesian framework---which inherently penalizes model complexity via Occam's razor---reveals that the absolute amplitudes of these modulations are so small that they do not warrant the addition of sinusoidal parameters over a simple null model (constant + linear drift). This discrepancy underscores that frequentist significance alone is insufficient for claiming physical discoveries in sub-percent neutrino flux modulations; Bayesian model comparison must serve as the ultimate arbiter.

\subsection{Physical Interpretation and Constraints on Detected Periodicities}
The hierarchical segmentation and Bayesian assessment allow us to rigorously constrain the physical origins of the candidate periodicities. 

The $\sim$0.106 yr ($\sim$38.8 d) period, detected by GLS in the pre-2001 and SK-I data, initially appears intriguing as it does not align with simple harmonics of the $\sim$27-day synodic solar rotation. Hypothetically, if this signal were of genuine astrophysical origin, it could be interpreted through the lens of resonant spin-flavor precession (RSFP) in the solar tachocline, where complex magnetic field geometries or latitudinal differential rotation could produce beat frequencies or sidebands distinct from the surface rotation rate. Alternatively, it might relate to a sub-harmonic of the quasi-biennial oscillation (QBO) observed in helioseismic data. 

\textit{However, we must exercise extreme caution regarding this interpretation.} As demonstrated in Section~\ref{sec:analysis_results}, the cross-method Bayesian consensus for this signal yields only \textit{weak evidence} ($\ln \mathrm{B} \approx 0.4$), and crucially, the signal is entirely absent in the SK-IV phase---the longest, highest-statistics, and most systematically stable dataset. This phase-dependent disappearance strongly suggests that the $\sim$38.8-day periodicity is not a persistent solar phenomenon, but rather a transient feature linked to the specific hardware configuration of SK-I, the solar cycle 23 rising phase, or a statistical fluctuation amplified by the lower event rates of the early years. 

Similarly, the $\sim$24.3 d signal detected in the post-2001 raw flux is decisively rejected by the Bayesian framework ($\ln \mathrm{B} \ll -5$) and vanishes upon $D^2$ distance correction. Its confinement to the raw flux of specific post-2001 epochs confirms its origin as a residual seasonal systematic or an instrumental alias, rather than a solar rotation harmonic.

Regarding the long-term solar cycle, the absence of any statistically significant $\sim$11-year modulation in the corrected flux is fully consistent with Standard Solar Model (SSM) predictions. Given the steep temperature dependence of the $^8$B production rate ($\Phi \propto T_c^{25}$), the helioseismically constrained core temperature variations over the solar cycle ($<0.1\%$) are far below the detection threshold of current experiments. Using the RMS scatter of the SK-IV modified flux ($\approx 0.015\times10^6$ cm$^{-2}$ s$^{-1}$) and the 2968-day baseline, we establish a stringent 95\% C.L. upper limit on the amplitude of any 11-year sinusoidal modulation of $<0.004\times10^6$ cm$^{-2}$ s$^{-1}$ ($\approx 0.2\%$ of the mean flux). This null result also aligns with recent theoretical evaluations demonstrating that second-order neutrino flux fluctuations induced by solar gravity (g) modes remain well below current experimental sensitivities~\cite{Hatta:2026gmode}.

\subsection{Systematic Uncertainties and Detector Effects}
The phase-resolved analysis explicitly isolates detector-specific artifacts from potential astrophysical signals. The strong dependence of the detected periods on the operational phase underscores the profound impact of hardware evolution:
\begin{itemize}
\item \textbf{SK-I to SK-II Transition:} The shift from the $\sim$38.8-day signal in SK-I to a $\sim$20.7-day signal in SK-II is highly indicative of instrumental origins. The SK-II phase, operating with a reduced PMT count (5,182 tubes) and a higher energy threshold, experienced altered spallation muon veto logic and trigger dead-times. The $\sim$20.7-day period is suspiciously close to the characteristic duty cycle of specific veto patterns or calibration runs deployed during the post-accident recovery phase, rather than a solar magnetic timescale.
\item \textbf{SK-III Window Function Artifacts:} The shortest baseline (548.5 days) inherently limits frequency resolution and exacerbates spectral leakage. The semi-annual ($\sim$203-day) and annual peaks observed in SK-III are classic signatures of window-function aliasing and imperfect subtraction of seasonal water transparency variations, which are known to exhibit $\sim$0.5--1 year periodicities.
\item \textbf{Directionality as an Intrinsic Safeguard:} Crucially, the directional reconstruction capability of water Cherenkov detectors provides a powerful, model-independent safeguard against non-solar backgrounds. By strictly selecting events pointing back to the Sun, Super-Kamiokande inherently suppresses isotropic backgrounds, including atmospheric neutrinos and potential off-axis signals from exotic scenarios (e.g., hypothetical dark matter annihilation scenarios proposed to explain anomalies in other experimental channels). The phase-by-phase stability analysis confirms that no known detector-specific alias consistently mimics a persistent astrophysical signal across the independent SK-I--IV configurations.
\end{itemize}

\subsection{Limitations and Future Prospects}
The current analysis is fundamentally limited by the $\sim$20 events per day rate, the irregular sampling (which introduces complex window functions), and the 22-year baseline covering only two solar cycles. However, the methodological framework established here provides a direct blueprint for the next generation of solar neutrino observatories.

Future detectors will not merely improve statistical precision; they will resolve the fundamental limitations of the SK dataset:
\begin{itemize}
\item \textbf{Hyper-Kamiokande (HK):} With a fiducial volume of $\sim$0.26 Mton, HK will increase the $^8$B event rate to $\sim$200 per day. More importantly, HK is designed for continuous, multi-decade operation without the catastrophic hardware interruptions (like the SK-I/II transition) that plague the current dataset. This continuity will virtually eliminate the spectral leakage and window-function artifacts that complicate the SK periodograms, enabling robust searches for sub-0.1\% modulations at higher frequencies.
\item \textbf{JUNO and Liquid Xenon Detectors:} The Jiangmen Underground Neutrino Observatory (JUNO) will probe the $pep$ and CNO neutrino channels via liquid scintillation, while next-generation dark matter detectors (e.g., DARWIN/XLZD) will observe the $^8$B ``neutrino fog'' via coherent elastic neutrino-nucleus scattering (CE$\nu$NS)~\cite{Zhuang:2024dm}. These entirely independent detection channels, possessing different systematic error profiles and energy thresholds, will allow for cross-correlation studies that can definitively separate solar-intrinsic modulations from detector-specific artifacts.
\end{itemize}

We strongly recommend that future multi-messenger and multi-detector solar neutrino analyses adopt the rigorous, multi-metric significance reporting framework (combining analytical/bootstrap FAPs with Bayesian MCMC model comparison) established in this work. As the field transitions from the era of ``signal detection'' to ``precision solar astrophysics,'' the open-source pipeline released alongside this paper~\footnote{\url{https://github.com/renlliang3/sk-neutrino-periodicity}} will serve as a standardized benchmark for evaluating temporal modulations in the high-statistics datasets of the 2030s.

\section{Conclusions}
\label{sec:conclusions}
We have presented a comprehensive, multi-method periodogram analysis of the complete 22-year Super-Kamiokande solar $^{8}\mathrm{B}$ neutrino dataset (1996--2018). By systematically comparing nine independent algorithms across classical, generalized, box-fitting, phase-based, and information-theoretic paradigms, and employing a hierarchical temporal segmentation strategy (Full, Pre/Post-July 2001, and SK-I--IV), we have rigorously disentangled potential astrophysical modulations from detector-specific systematics and orbital residuals. Our analysis establishes a new methodological benchmark for time-series periodicity searches in low-statistics, heteroscedastic neutrino data. 

The main conclusions of this work are as follows:

\begin{enumerate}
\item \textbf{Methodological Consistency and the Value of Negative Results:}
The Generalized Lomb-Scargle (GLS) method provides the highest statistical significance (FAP $< 0.001$) for the detected periodicities by properly accounting for heteroscedastic uncertainties, whereas the classical LS method yields marginal significance despite locating the correct frequencies. Crucially, multiple independent algorithms---including phase-dispersion (PDM1, MHAOV) and box-fitting (BLS) methods---consistently recover the same periodicities as GLS across various temporal segments (Tables~\ref{tab:period_temporal_raw_mod} and \ref{tab:period_sk_raw_mod}). This cross-method consensus confirms the robustness of the extracted signals. Furthermore, the non-detections by information-theoretic (QME, QMICS, QMIEU) and box-fitting (BLS) methods constitute highly informative \textit{negative results}. They confirm that the intrinsic temporal modulations of the solar $^8$B flux are overwhelmingly well-approximated by smooth, sinusoidal waveforms, placing stringent upper limits ($<2\%$ of the mean flux) on short-duration, non-sinusoidal transit-like depressions. Finally, the stark contrast between frequentist FAPs and Bayesian Bayes factors ($\ln \mathrm{B}$) highlights a critical paradigm: frequentist significance alone is insufficient for claiming physical discoveries in sub-percent flux modulations; Bayesian model comparison, which penalizes complexity for low-amplitude signals, must serve as the ultimate arbiter.

\item \textbf{Constraints on Long-Term Solar Modulations:} 
No evidence is found for an $\sim$11-year solar cycle modulation in the distance-corrected flux. Using the RMS scatter of the SK-IV modified flux ($\approx 0.015\times10^6$ cm$^{-2}$ s$^{-1}$) as a reference, we establish a 95\% C.L. upper limit of $<0.004\times10^6$ cm$^{-2}$ s$^{-1}$ on the amplitude of any 11-year sinusoidal modulation, corresponding to $\approx 0.2\%$ of the mean SK-IV flux.). This null result is fully consistent with Standard Solar Model predictions and recent theoretical evaluations of solar gravity (g) modes, confirming that core temperature variations and second-order neutrino flux fluctuations remain well below current experimental sensitivities.

\item \textbf{Re-evaluation of Short-Period Candidates:} 
The hierarchical segmentation reveals that previously debated short-period signals are transient or systematic in nature. The $\sim$0.106 yr ($\sim$38.8 d) periodicity, while exhibiting frequentist significance in the pre-2001 and SK-I data, yields only \textit{weak evidence} ($\ln \mathrm{B} \approx 0.4$) in the Bayesian framework and is entirely absent in the longest, most systematically stable SK-IV phase. We therefore interpret this signal not as a persistent astrophysical phenomenon (such as a stable RSFP or rotation harmonic), but as a transient feature of the early low-statistics era or a solar-cycle-dependent amplitude modulation. Similarly, the $\sim$24.3 d and $\sim$20.7 d signals detected in specific post-2001 raw flux segments are decisively rejected by the Bayesian consensus ($\ln \mathrm{B} \ll -5$) and vanish upon distance correction, conclusively identifying them as residual seasonal systematics or detector-specific aliases rather than solar rotation harmonics.

\item \textbf{Roadmap for Next-Generation Observatories:} 
The systematic artifacts and window-function aliasing exposed in this 22-year dataset underscore the critical importance of continuous, uninterrupted operation for future solar neutrino observatories. Hyper-Kamiokande (HK) will not only increase the event rate by an order of magnitude, enabling sub-0.1\% modulation searches, but its anticipated multi-decade continuous operation will virtually eliminate the spectral leakage issues that complicate the SK periodograms. Complementarily, JUNO and next-generation liquid xenon detectors (e.g., DARWIN/XLZD) will provide independent detection channels with distinct systematic profiles, allowing for definitive cross-correlation studies to separate solar-intrinsic modulations from local detector effects.

\end{enumerate}

To facilitate reproducibility and establish a standardized framework for the upcoming era of precision solar neutrino astrophysics, the complete open-source analysis pipeline---encompassing the preprocessing, nine periodogram implementations, bootstrap-EVT significance estimators, and Bayesian MCMC routines---is publicly released.\footnote{\url{https://github.com/renlliang3/sk-neutrino-periodicity}}.The multi-metric significance reporting framework established in this work is strongly recommended for all future neutrino time-series publications to ensure rigorous, cross-experiment comparability.

\acknowledgments

We acknowledge the Super-Kamiokande Collaboration for providing the open-access solar neutrino dataset. This work utilized computational resources from the High-Performance Computing Center at Anhui Science and Technology University. Funding was provided by the National Natural Science Foundation of China under grants No. 12305056, and the Anhui Science and Technology University's Key Discipline Construction Fund (XK-XJGY002). We thank Dr. Pablo Huijse for developing the P4J package and for helpful discussions on QMI implementation. We also thank the anonymous reviewers for constructive comments that improved the manuscript.

The following packages were used for this work: Astropy \cite{astropy:2013,astropy:2022,astropy:2018}, NumPy \cite{harris2020array}, SciPy \cite{2020SciPy-NMeth}, Matplotlib \cite{Hunter:2007}, P4J \cite{Huijse:2018Ap}, emcee \cite{Foreman:2013PA}.

\appendix
\section{Detailed Periodogram Results for All Temporal Segments}
\label{app:periodograms}

\begin{figure}[tbp]
\centering
\includegraphics[width=15.5cm]{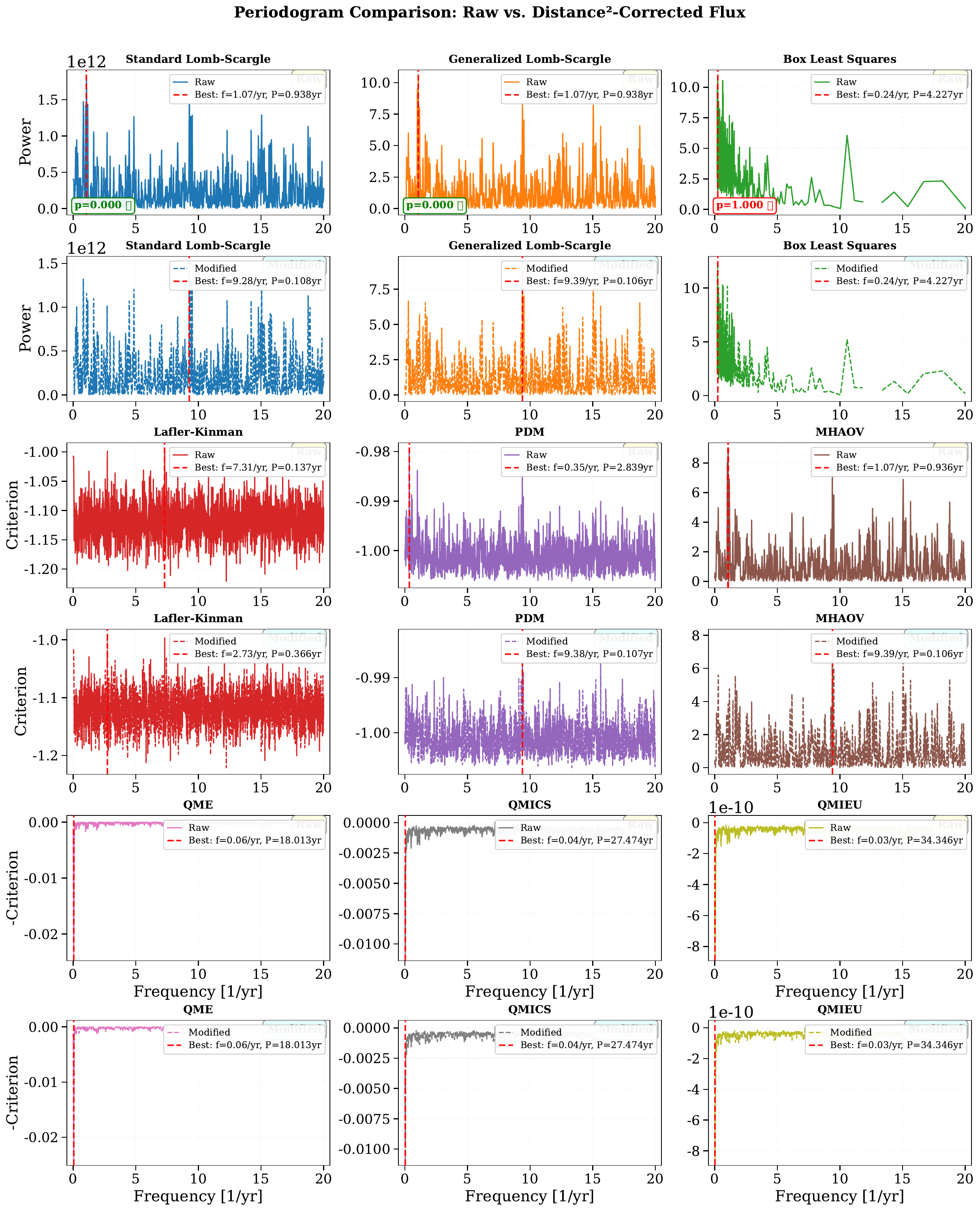}
\caption{Comprehensive periodogram analysis of the complete 22-year Super-Kamiokande solar $^{8}\mathrm{B}$ neutrino dataset (1996--2018) using nine independent algorithms. The multi-panel layout displays periodograms for raw flux measurements (top rows) and distance-corrected (modified) flux (bottom rows) after applying the $D^2$ correction (Eq.~\ref{eq:d2_correction}). In the raw flux, GLS robustly identifies the dominant $\sim$1-yr period ($f \approx 1.06$ yr$^{-1}$) driven by Earth's orbital eccentricity, while LS yields marginal significance due to unmodeled heteroscedasticity. Following the $D^2$ correction, the 1-yr peak is effectively suppressed in the modified flux, and GLS recovers a significant $\sim$38.8-day ($0.106$ yr) period (FAP $< 0.001$), a signal also independently recovered by MHAOV and PDM1 at consistent frequencies. Notably, the information-theoretic algorithms (QME, QMICS, QMIEU) exhibit severe degeneracies in this specific dataset, yielding flat or grid-boundary power spectra, indicating their lower sensitivity to the sinusoidal modulations present compared to variance-based methods like GLS.}
\label{fig:full_periodogram}
\end{figure}

\begin{figure}[tbp]
\centering
\includegraphics[width=15.5cm]{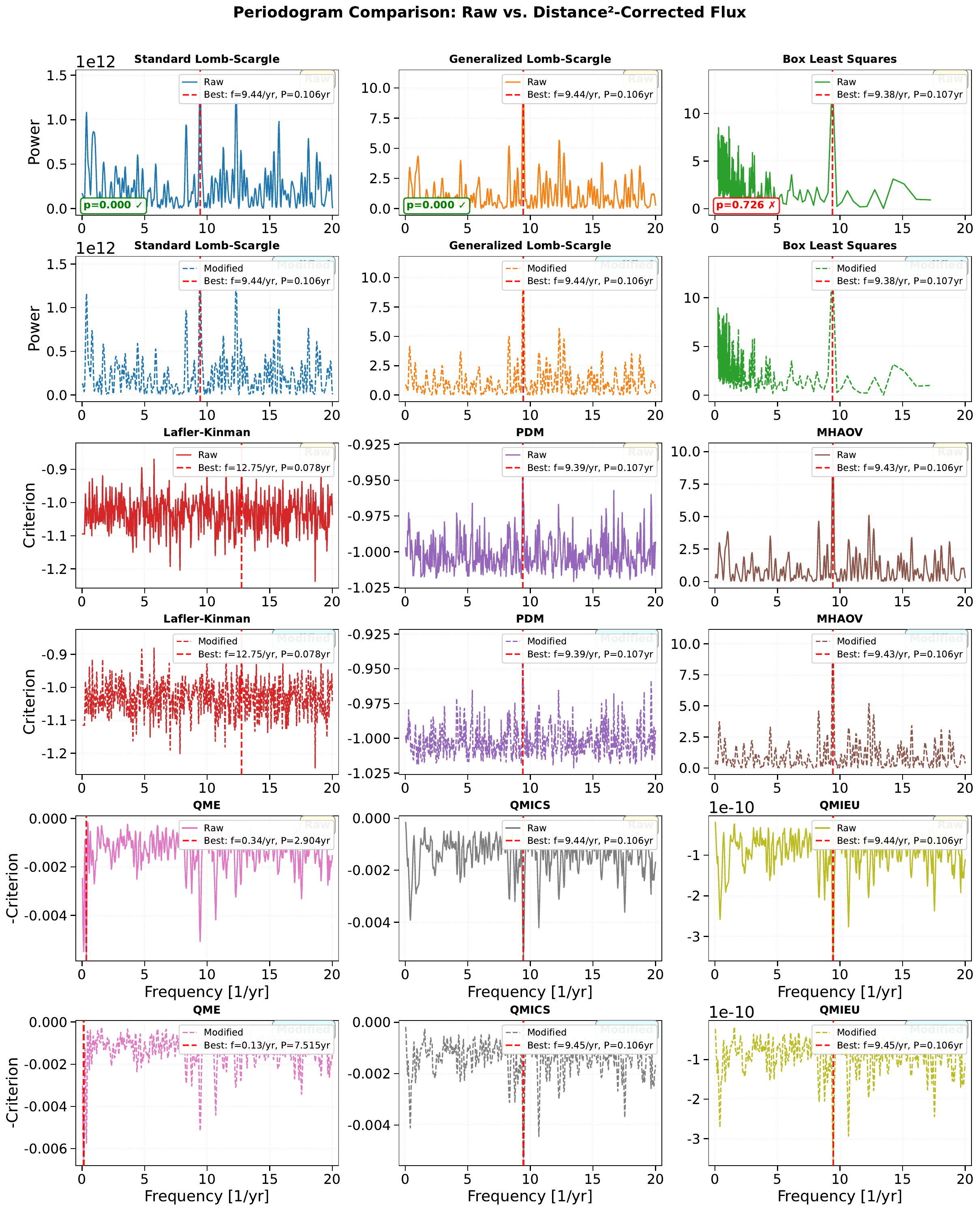}
\caption{Periodogram analysis of the pre-July 2001 data (SK-I era, 1495.7 live days). Layout and algorithm arrangement are identical to Figure~\ref{fig:full_periodogram}.}
\label{fig:pre2001_periodogram}
\end{figure}

\begin{figure}[tbp]
\centering
\includegraphics[width=15.5cm]{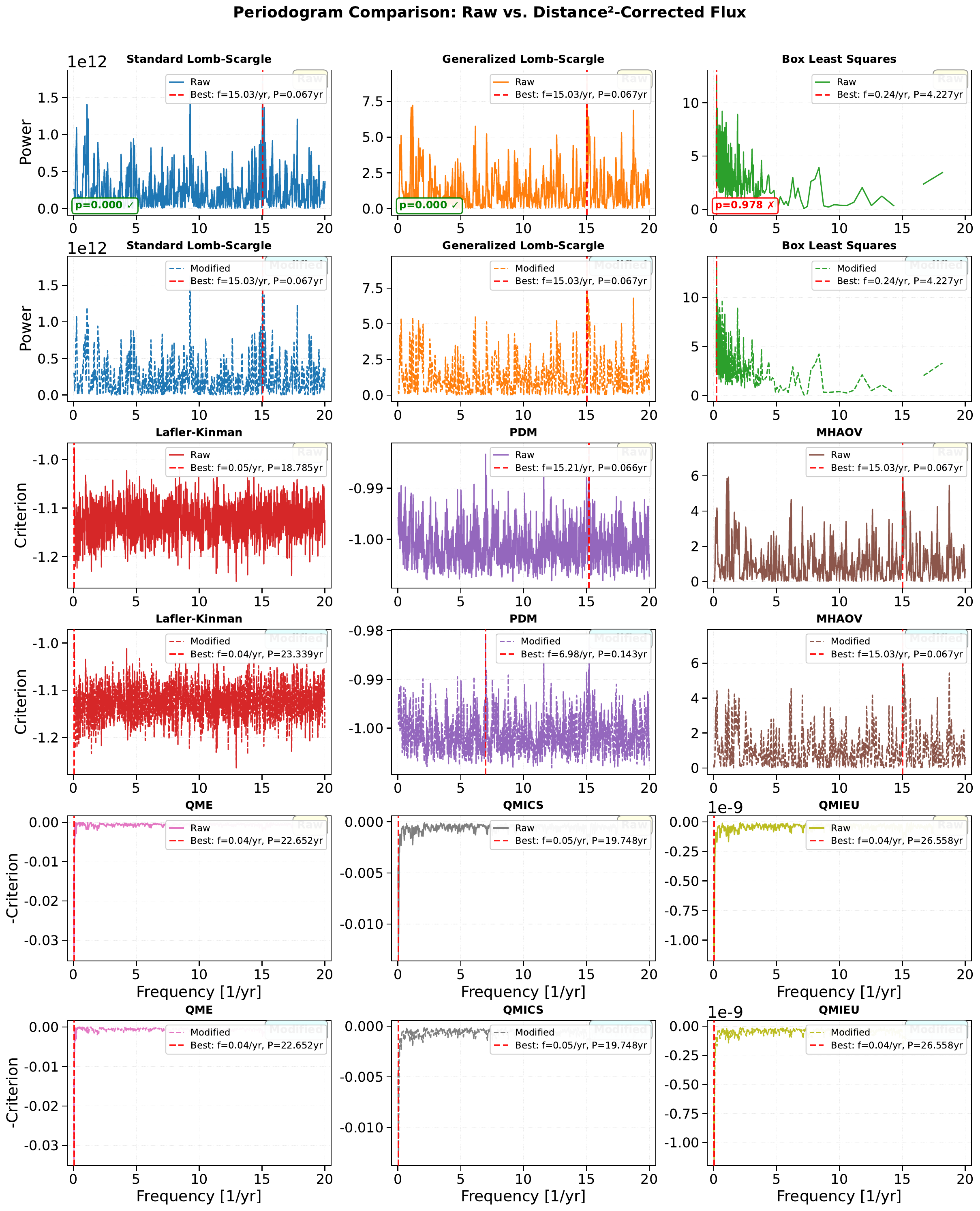}
\caption{Periodogram analysis of the post-July 2001 data (combined SK-II--IV, 4308.1 live days). Layout identical to Figure~\ref{fig:full_periodogram}.}
\label{fig:post2001_periodogram}
\end{figure}

\begin{figure}[tbp]
\centering
\includegraphics[width=15.5cm]{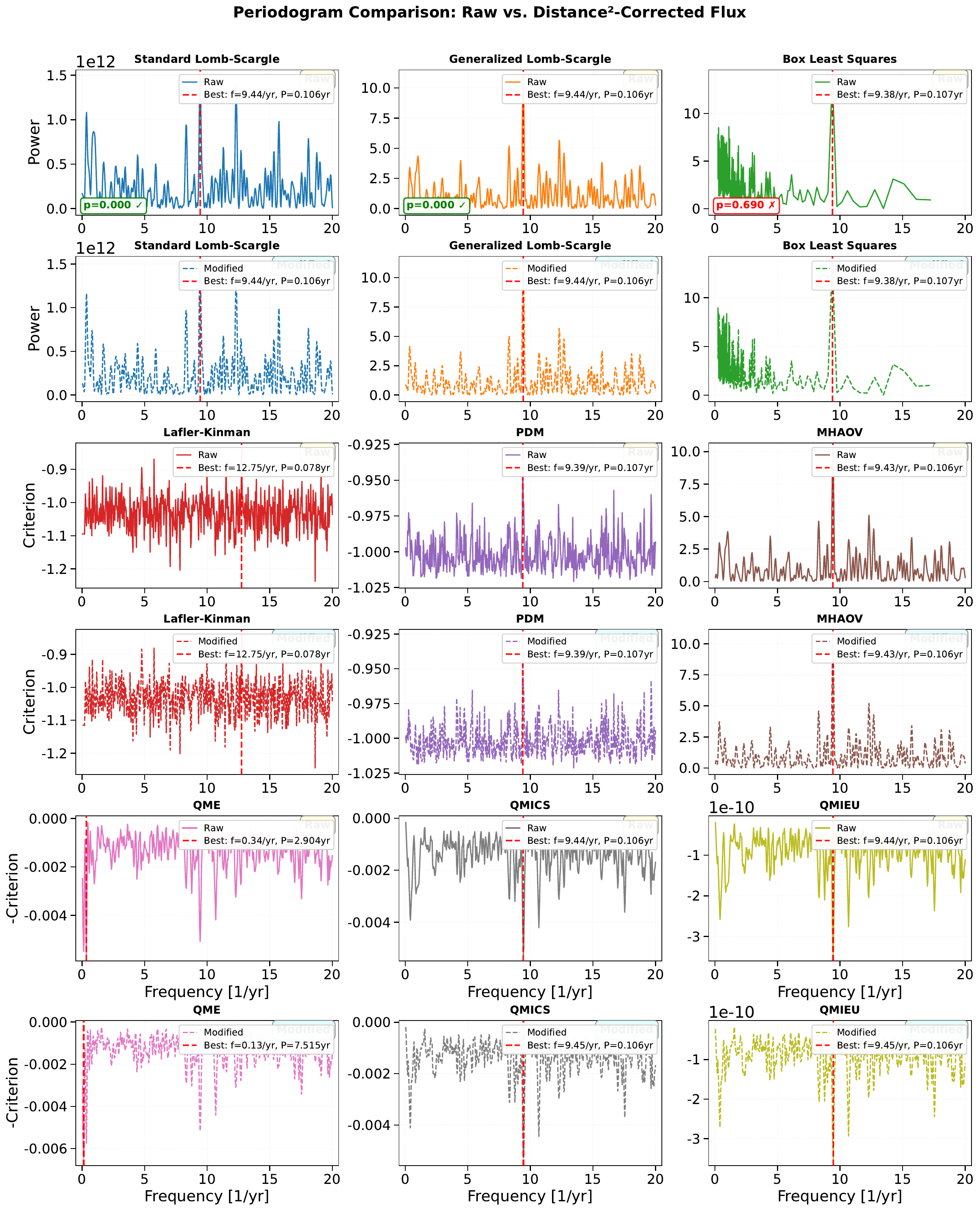}
\caption{Periodogram analysis of the SK-I phase (1996--2001, 1495.7 live days). Layout identical to Figure~\ref{fig:full_periodogram}.}
\label{fig:ski_periodogram}
\end{figure}

\begin{figure}[tbp]
\centering
\includegraphics[width=15.5cm]{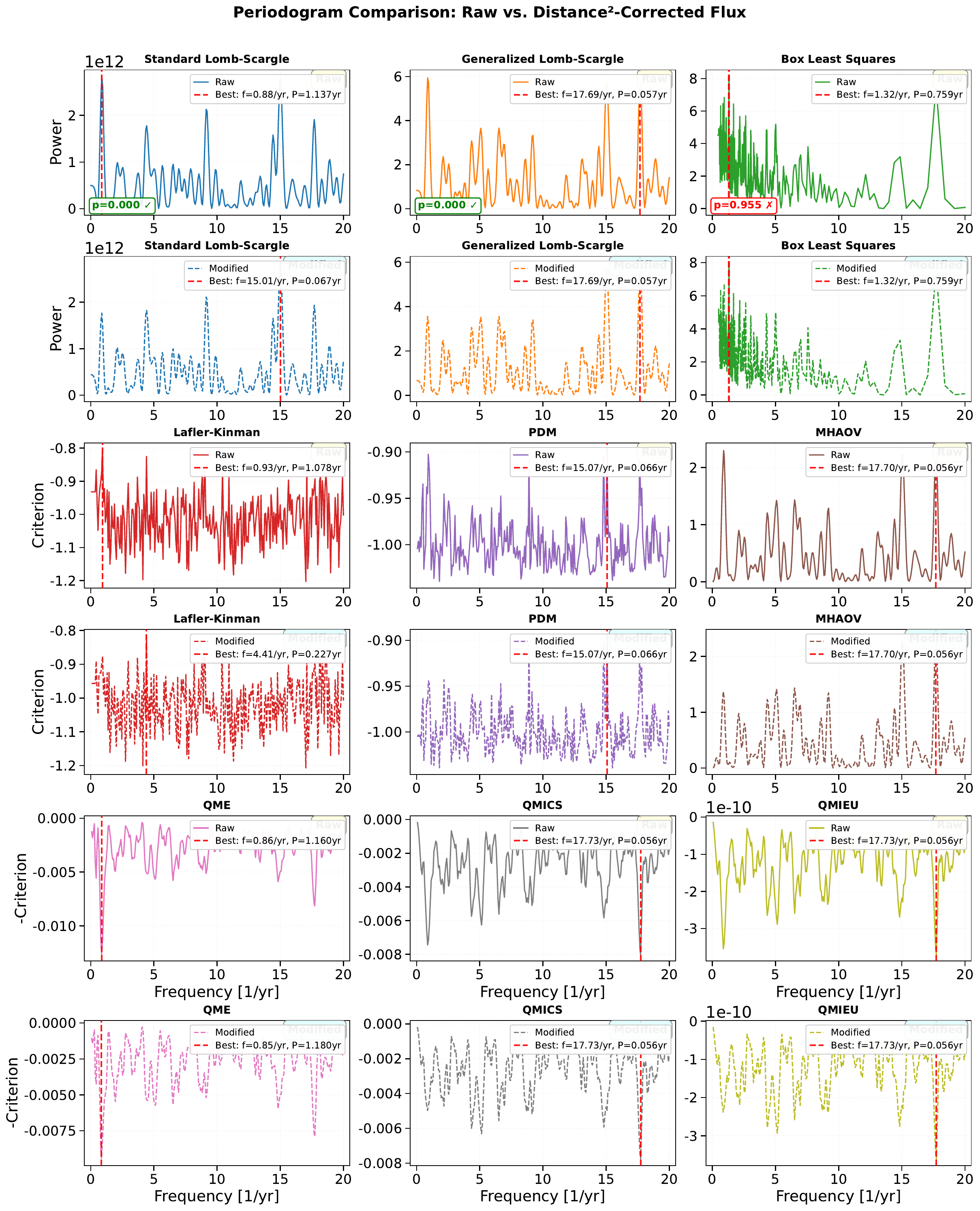}
\caption{Periodogram analysis of the SK-II phase (2002--2005, 791.9 live days). Layout identical to Figure~\ref{fig:full_periodogram}.}
\label{fig:skii_periodogram}
\end{figure}

\begin{figure}[tbp]
\centering
\includegraphics[width=15.5cm]{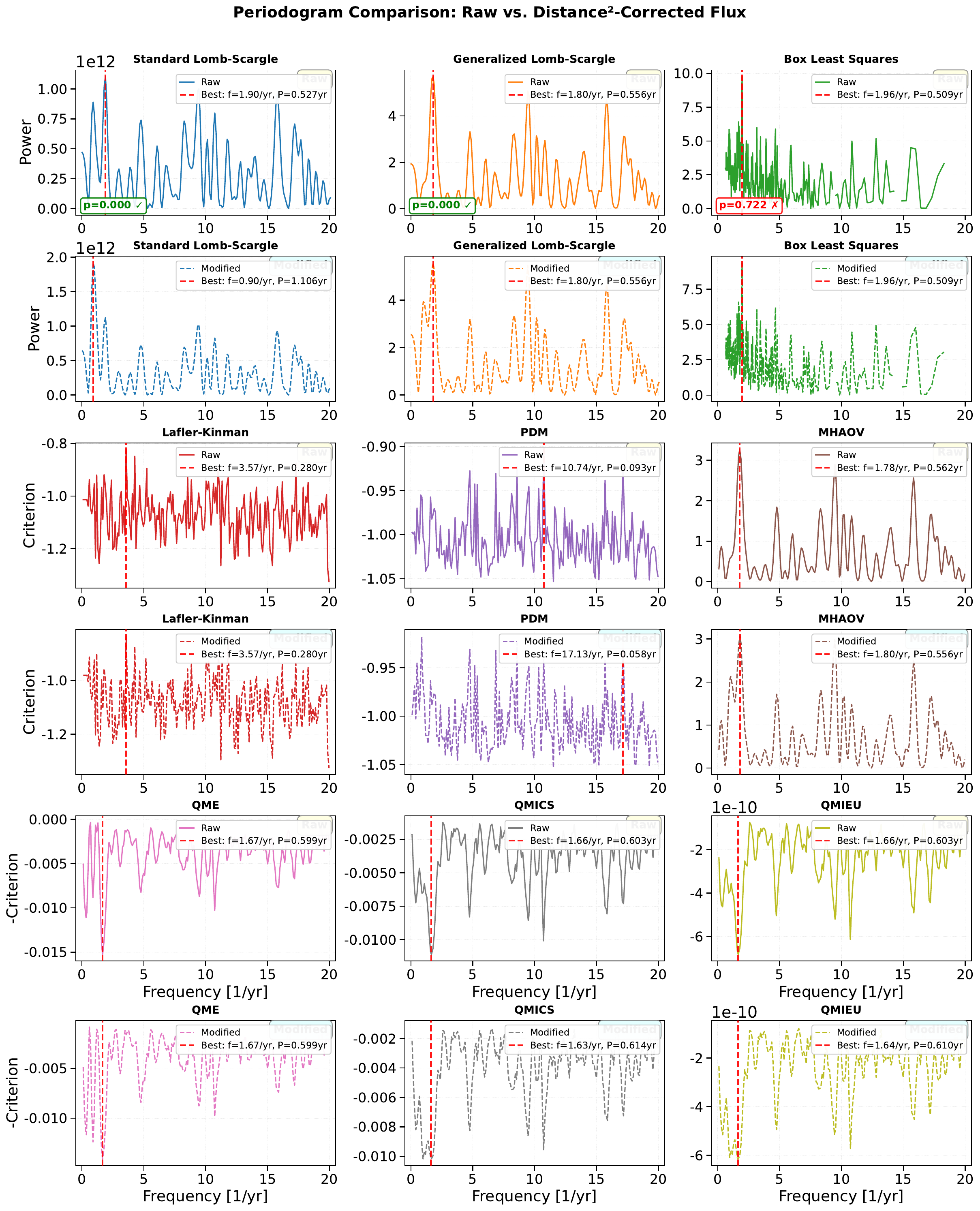}
\caption{Periodogram analysis of the SK-III phase (2006--2008, 548.5 live days). Layout identical to Figure~\ref{fig:full_periodogram}.}
\label{fig:skiii_periodogram}
\end{figure}

\begin{figure}[tbp]
\centering
\includegraphics[width=15.5cm]{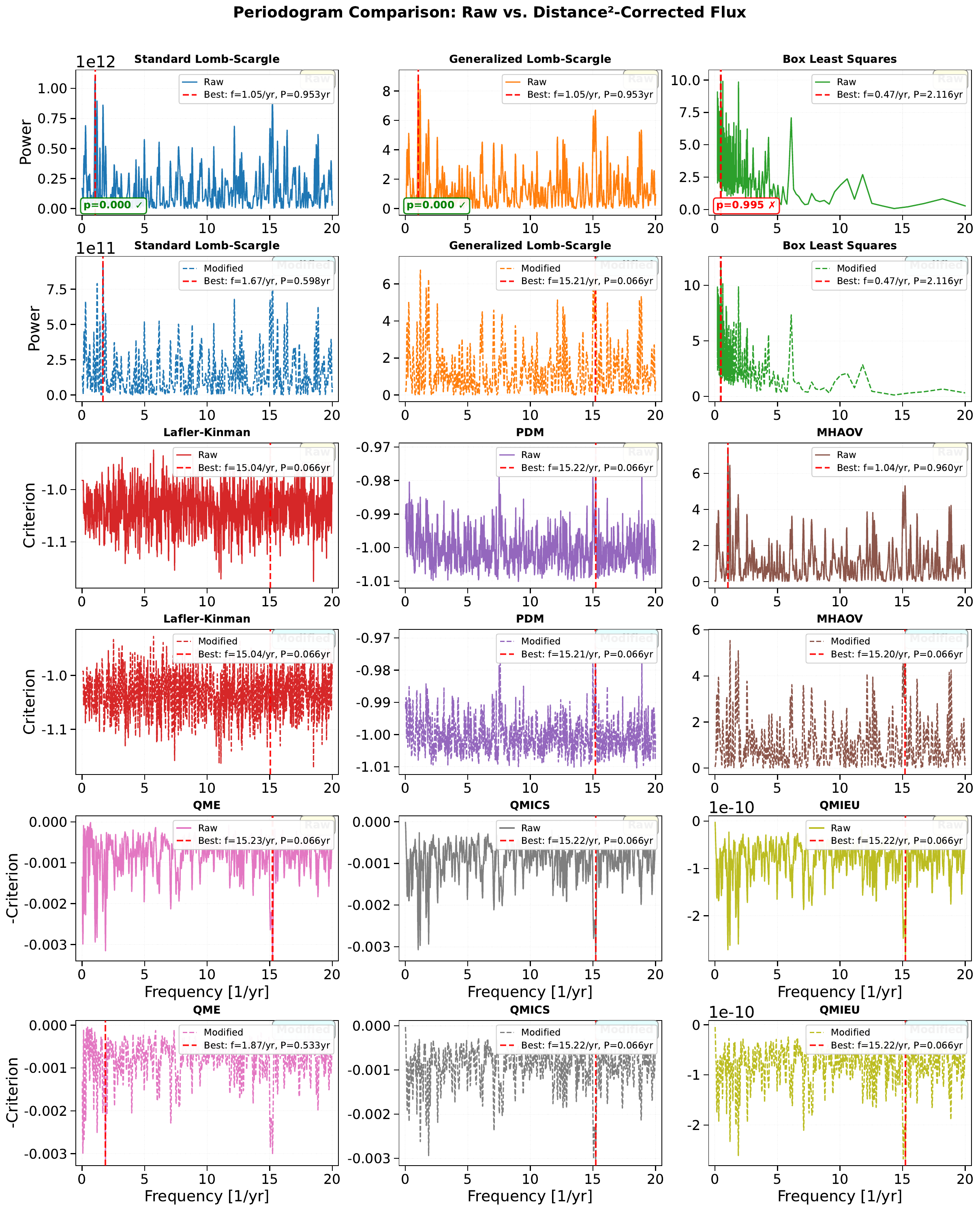}
\caption{Periodogram analysis of the SK-IV phase (2008--2018, 2967.7 live days). Layout identical to Figure~\ref{fig:full_periodogram}.}
\label{fig:skiv_periodogram}
\end{figure}

This appendix presents the comprehensive periodogram analyses for each temporal segment evaluated in Section~\ref{sec:analysis_results}. The detailed periodogram comparisons for all nine algorithms are displayed in Figures~\ref{fig:full_periodogram}--\ref{fig:skiv_periodogram}, which cover the full 22-year baseline (Figure~\ref{fig:full_periodogram}), the pre- and post-July 2001 epochs (Figures~\ref{fig:pre2001_periodogram} and \ref{fig:post2001_periodogram}), and the four individual SK-I--IV operational phases (Figures~\ref{fig:ski_periodogram}--\ref{fig:skiv_periodogram}). These figures display the results of the nine independent period-finding algorithms applied to both raw and distance-corrected (modified) flux measurements. Each figure follows a standardized multi-panel layout, facilitating direct visual comparison of algorithmic performance and sensitivity.

\section{MCMC Posterior Corner Plots}
\label{app:corner}
\begin{figure}[tbp]
\centering
\includegraphics[width=0.75\textwidth]{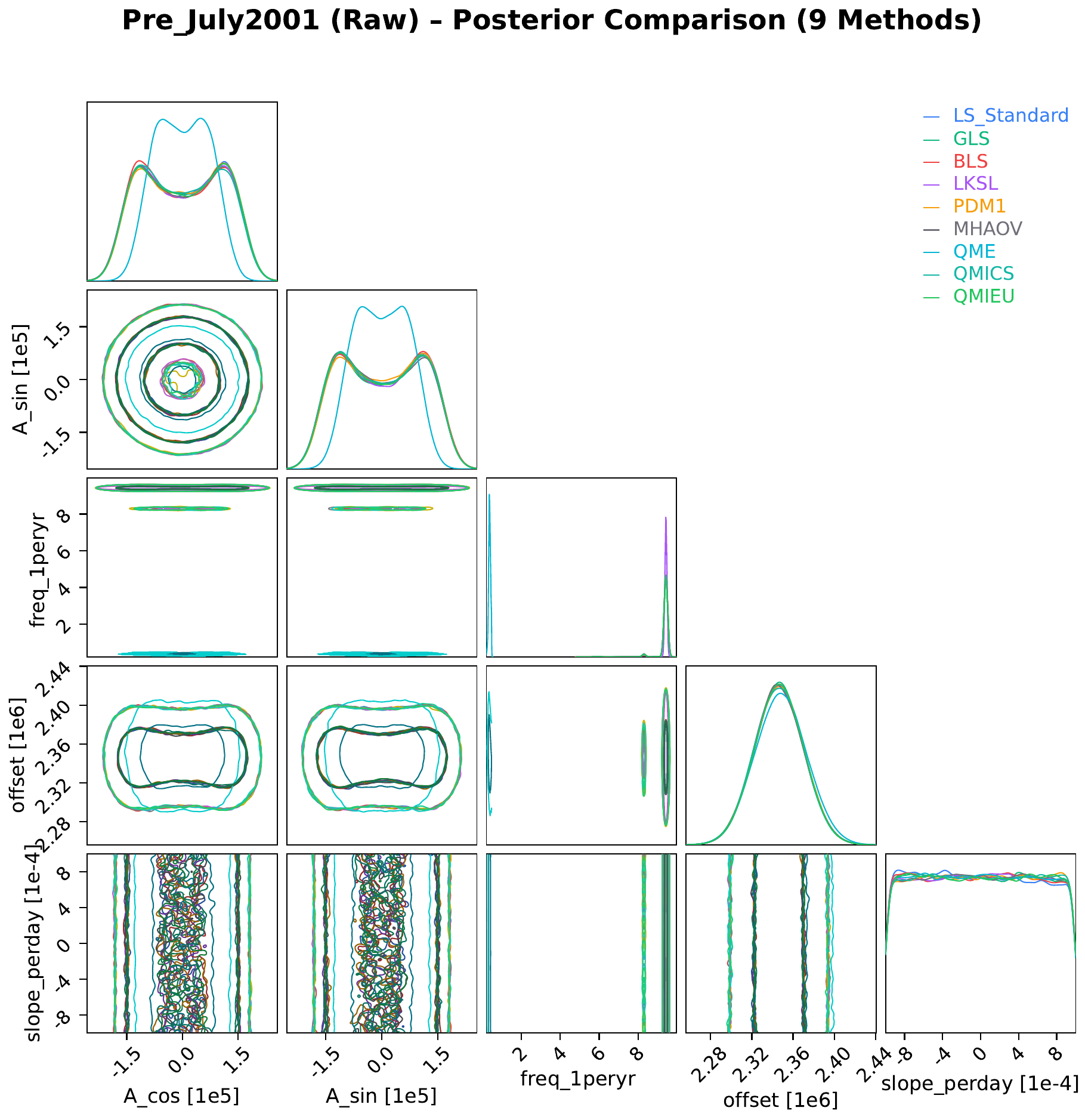} \\
\vspace{0.5em}
\includegraphics[width=0.75\textwidth]{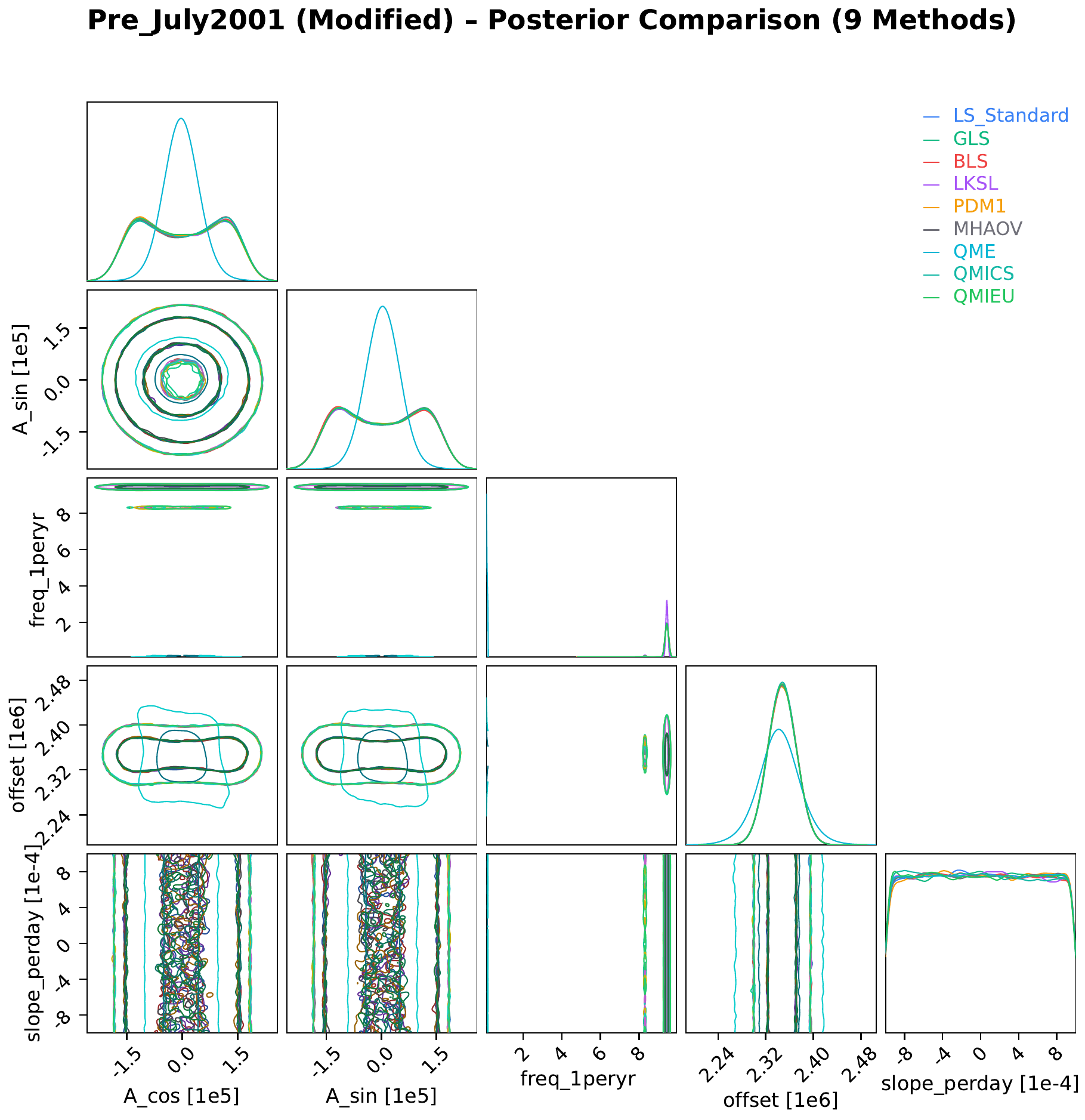}
\caption{Corner plots for the pre-July 2001 segment (SK-I era). \textbf{Top:} Raw flux; \textbf{Bottom:} Modified flux.}
\label{fig:corner_pre}
\end{figure}

This appendix presents a representative corner plot (pairwise joint posterior distributions) derived from the Bayesian MCMC sinusoidal fits. Here, we select the results for the pre-July 2001 segment (SK-I era) as a representative example, as it contains the most statistically notable periodic signal ($\sim$38.8-day) identified in this study. The fitted parameter vector is $\theta = \{A, B, \nu, c, s\}$, comprising the cosine amplitude $A$, sine amplitude $B$, frequency $\nu$ (in yr$^{-1}$), constant offset $c$, and linear drift slope $s$ (in day$^{-1}$). As shown in Figure~\ref{fig:corner_pre}, the left panel corresponds to the raw flux, and the right panel corresponds to the distance-corrected (modified) flux. Corner plots for the remaining temporal segments exhibit qualitatively similar posterior structures and are available in the public analysis repository.

\section{Best-fit Parameters from Bayesian MCMC Fits}
\label{app:bestfit}

To rigorously validate the periodicities identified by the nine periodogram algorithms and to quantify the corresponding parameter uncertainties, we perform Bayesian Markov Chain Monte Carlo (MCMC) sinusoidal fits for each temporal segment. The fitting procedure strictly adopts the time-binned formalism established by the Super-Kamiokande collaboration, as detailed in Section~\ref{subsec:bayesian}. Specifically, we fit the bin-averaged sinusoidal model (Eqs.~\ref{eq:prl_bin_avg_integral} and \ref{eq:prl_bin_avg_analytic}) to the data, utilizing the best-fit frequency derived from the periodogram peak as the initial guess for the MCMC sampler.

Table~\ref{tab:combined_bestfit_params} comprehensively summarizes the posterior evidence and the corresponding best-fit parameters for all evaluated segment--method combinations. For each fit, the table reports the statistical metrics used for model comparison: the best period $P_{\rm best}$, the logarithmic Bayes factor $\ln \mathrm{B}$ (computed via the Bayesian Information Criterion approximation as described in Section~\ref{subsec:bayesian}) relative to a null model consisting of a constant plus a linear drift, and the goodness-of-fit indicators (BIC and $\chi^2$). 

Additionally, the table lists the physical parameters of the sinusoidal model: the cosine and sine amplitudes ($A$ and $B$), the angular frequency $\nu$ (in yr$^{-1}$), the constant offset, and the linear slope (in day$^{-1}$). All quoted parameter values represent the medians of the marginalized posterior distributions. Asymmetric $1\sigma$ uncertainties (corresponding to the 16th and 84th percentiles) are omitted for brevity. Flux-related quantities ($A$, $B$, and offset) are given in units of $10^6\,\mathrm{cm}^{-2}\,\mathrm{s}^{-1}$. The interpretation of the logarithmic Bayes factor $\ln \mathrm{B}$ follows Jeffreys' scale, where positive values indicate evidence in favor of the periodic sinusoidal model, and negative values signify that the null model is preferred.

\begingroup
\small
\setlength{\tabcolsep}{4pt}
\begin{longtable}{@{} l c c c c c c c c c @{}}
\caption{Combined best-fit parameters and summary statistics (Scheme A: Raw \& Modified). Parameters ($A$, $B$, offset) in raw units; $\nu$ in yr$^{-1}$; slope in day$^{-1}$. Statistics: best period, logarithmic Bayes factor $\ln \mathrm{B}$ (BIC approximation, Eq.~3.27), BIC, and $\chi^2$. All values are posterior medians.}
\label{tab:combined_bestfit_params}\\
\toprule
\textbf{Method} & $A$ & $B$ & $\nu$ (yr$^{-1}$) & offset & slope & $P_{\rm best}$ (yr) & $\ln \mathrm{B}$ & \textbf{BIC} & $\chi^2$ \\
\midrule
\endfirsthead
\caption*{\textit{Continued} -- Combined summary (cont.)}\\
\toprule
\textbf{Method} & $A$ & $B$ & $\nu$ (yr$^{-1}$) & offset & slope & $P_{\rm best}$ (yr) & $\ln \mathrm{B}$ & \textbf{BIC} & $\chi^2$ \\
\midrule
\endhead
\midrule
\multicolumn{10}{r}{\footnotesize \textit{Continued on next page}} \\
\endfoot
\bottomrule
\endlastfoot

\multicolumn{10}{l}{\textbf{Full\_22years (Raw)}} \\
\midrule
LS\_Standard & 21606.10 & 12096.56 & 1.0695 & 2.34e+06 & -1.01e-04 & 0.9382 & 0.18 & 1620.8 & 1584.8 \\
GLS & 9239.00 & 24185.00 & 1.0674 & 2.34e+06 & -1.05e-04 & 0.9382 & 0.18 & 1620.8 & 1584.8 \\
BLS & 1278.66 & 3085.42 & 0.2650 & 2.34e+06 & 5.26e-05 & 4.2270 & -6.13 & 1633.4 & 1597.4 \\
LKSL & 1983.50 & 1647.26 & 8.0182 & 2.34e+06 & 7.83e-05 & 0.1368 & -2.20 & 1625.6 & 1589.6 \\
PDM1 & 2703.98 & 3622.60 & 0.2837 & 2.34e+06 & 1.32e-04 & 2.8388 & -6.13 & 1633.4 & 1597.4 \\
MHAOV & 37077.50 & 18105.49 & 1.0676 & 2.34e+06 & 1.61e-04 & 0.9358 & 0.17 & 1620.8 & 1584.8 \\
QME & 252.36 & 4159.16 & 0.0760 & 2.34e+06 & -9.61e-06 & 18.0135 & -10.31 & 1641.8 & 1605.8 \\
QMICS & -3352.63 & -13882.61 & 0.0503 & 2.35e+06 & 2.07e-05 & 27.4739 & -9.74 & 1640.6 & 1604.6 \\
QMIEU & -16121.75 & 9623.98 & 0.0417 & 2.34e+06 & 4.60e-05 & 34.3455 & -9.74 & 1640.6 & 1604.6 \\
\midrule
\multicolumn{10}{l}{\textbf{Full\_22years (Modified)}} \\
\midrule
LS\_Standard & -533.00 & 3481.02 & 9.3767 & 2.33e+06 & -1.09e-05 & 0.1078 & -2.03 & 1610.6 & 1574.6 \\
GLS & 6517.38 & 2199.80 & 9.3888 & 2.34e+06 & -6.30e-05 & 0.1064 & -2.03 & 1610.6 & 1574.6 \\
BLS & 6024.53 & -991.13 & 0.2762 & 2.34e+06 & 2.75e-05 & 4.2270 & -5.77 & 1618.0 & 1582.0 \\
LKSL & -10656.42 & 124.29 & 2.0000 & 2.34e+06 & 3.32e-05 & 0.3661 & -5.13 & 1616.9 & 1580.9 \\
PDM1 & -991.98 & 5818.05 & 9.3845 & 2.34e+06 & -1.23e-04 & 0.1066 & -2.03 & 1610.7 & 1574.7 \\
MHAOV & 420.54 & 1474.56 & 8.7104 & 2.33e+06 & 2.75e-04 & 0.1065 & -2.03 & 1610.6 & 1574.6 \\
QME & -811.25 & 3980.37 & 0.0750 & 2.34e+06 & -2.55e-05 & 18.0135 & -10.83 & 1628.1 & 1592.1 \\
QMICS & -4328.95 & -10329.18 & 0.0499 & 2.34e+06 & 5.52e-05 & 27.4739 & -10.02 & 1626.5 & 1590.5 \\
QMIEU & -12840.67 & 10307.78 & 0.0407 & 2.34e+06 & 7.05e-05 & 34.3455 & -10.02 & 1626.5 & 1590.5 \\
\midrule

\multicolumn{10}{l}{\textbf{Pre\_July2001 (Raw)}} \\
\midrule
LS\_Standard & -1659.85 & 14200.75 & 9.4177 & 2.34e+06 & 2.15e-05 & 0.1059 & 0.11 & 415.9 & 386.5 \\
GLS & 5230.97 & 25190.28 & 9.4067 & 2.35e+06 & -2.41e-04 & 0.1059 & 0.11 & 415.9 & 386.5 \\
BLS & 4126.10 & 3051.55 & 9.4099 & 2.34e+06 & 1.42e-04 & 0.1066 & 0.11 & 415.9 & 386.5 \\
LKSL & -78.46 & -17919.31 & 9.3989 & 2.35e+06 & 2.95e-05 & 0.0784 & 0.11 & 416.0 & 386.6 \\
PDM1 & -10990.88 & -6527.71 & 9.4154 & 2.35e+06 & 1.06e-04 & 0.1065 & 0.11 & 416.0 & 386.6 \\
MHAOV & 1175.28 & -24244.44 & 9.4068 & 2.34e+06 & -2.00e-04 & 0.1061 & 0.11 & 416.0 & 386.6 \\
QME & -13551.81 & -6798.85 & 0.3734 & 2.35e+06 & -7.56e-05 & 2.9036 & -5.56 & 427.3 & 397.9 \\
QMICS & 6545.06 & -4186.87 & 9.4179 & 2.35e+06 & -7.99e-05 & 0.1059 & 0.11 & 415.9 & 386.5 \\
QMIEU & 6104.99 & -19943.72 & 9.3922 & 2.35e+06 & 1.00e-04 & 0.1059 & 0.11 & 415.9 & 386.5 \\
\midrule
\multicolumn{10}{l}{\textbf{Pre\_July2001 (Modified)}} \\
\midrule
LS\_Standard & 22462.19 & 6977.71 & 9.4172 & 2.35e+06 & 8.21e-05 & 0.1059 & 0.40 & 408.4 & 379.0 \\
GLS & 26733.75 & -30346.45 & 9.4149 & 2.35e+06 & -1.19e-05 & 0.1059 & 0.40 & 408.4 & 379.0 \\
BLS & -19599.09 & 7500.86 & 9.4121 & 2.35e+06 & -3.37e-05 & 0.1066 & 0.40 & 408.4 & 379.0 \\
LKSL & -28789.74 & 46351.77 & 9.4171 & 2.34e+06 & 2.00e-05 & 0.0784 & 0.40 & 408.4 & 379.0 \\
PDM1 & 13145.88 & -4988.09 & 9.4119 & 2.35e+06 & 1.16e-04 & 0.1065 & 0.40 & 408.4 & 379.0 \\
MHAOV & 19149.31 & 14341.79 & 9.4070 & 2.35e+06 & 3.50e-05 & 0.1061 & 0.40 & 408.4 & 379.0 \\
QME & -4920.36 & 6433.08 & 0.1288 & 2.34e+06 & 6.86e-05 & 7.5151 & -8.98 & 427.2 & 397.8 \\
QMICS & 5944.84 & -14561.01 & 9.4179 & 2.34e+06 & 2.94e-05 & 0.1058 & 0.40 & 408.4 & 379.0 \\
QMIEU & 24.10 & 11875.10 & 9.4179 & 2.35e+06 & -7.47e-05 & 0.1058 & 0.40 & 408.5 & 379.1 \\
\midrule

\multicolumn{10}{l}{\textbf{Post\_July2001 (Raw)}} \\
\midrule
LS\_Standard & 2852.36 & 2875.15 & 9.4125 & 2.34e+06 & -1.02e-04 & 0.0665 & -7.28 & 1228.3 & 1193.8 \\
GLS & 3003.97 & 7475.79 & 9.4109 & 2.33e+06 & 2.03e-05 & 0.0665 & -7.29 & 1228.2 & 1193.8 \\
BLS & -5511.45 & 2027.85 & 0.2444 & 2.33e+06 & -1.68e-05 & 4.2270 & -6.15 & 1225.9 & 1191.5 \\
LKSL & 2466.78 & 3406.16 & 0.0734 & 2.34e+06 & 5.66e-05 & 18.7849 & -10.56 & 1234.8 & 1200.3 \\
PDM1 & -359.42 & -3248.37 & 9.4393 & 2.34e+06 & 6.32e-05 & 0.0658 & -7.29 & 1228.3 & 1193.8 \\
MHAOV & -5009.54 & -5703.72 & 9.4120 & 2.34e+06 & 3.25e-05 & 0.0665 & -7.29 & 1228.2 & 1193.7 \\
QME & 3345.78 & -2801.30 & 0.0600 & 2.34e+06 & -3.36e-05 & 22.6523 & -10.38 & 1234.4 & 1199.9 \\
QMICS & 1768.16 & 2116.30 & 0.0704 & 2.34e+06 & 4.68e-05 & 19.7482 & -10.51 & 1234.7 & 1200.2 \\
QMIEU & -836.15 & -8401.99 & 0.0508 & 2.35e+06 & -1.63e-05 & 26.5579 & -10.29 & 1234.2 & 1199.7 \\
\midrule
\multicolumn{10}{l}{\textbf{Post\_July2001 (Modified)}} \\
\midrule
LS\_Standard & 10979.17 & 4555.12 & 9.4077 & 2.33e+06 & -6.20e-05 & 0.0665 & -7.07 & 1220.1 & 1185.7 \\
GLS & 1336.42 & 2081.41 & 9.4092 & 2.33e+06 & 1.15e-04 & 0.0665 & -7.08 & 1220.1 & 1185.6 \\
BLS & -6804.07 & -1526.44 & 0.2486 & 2.33e+06 & 2.94e-05 & 4.2270 & -6.17 & 1218.2 & 1183.8 \\
LKSL & 4397.51 & -1514.50 & 0.0568 & 2.34e+06 & 2.92e-05 & 23.3388 & -10.64 & 1227.2 & 1192.7 \\
PDM1 & -4232.07 & 4152.99 & 6.1747 & 2.33e+06 & -9.82e-05 & 0.1434 & -5.96 & 1217.9 & 1183.4 \\
MHAOV & -6276.38 & 2471.18 & 9.3928 & 2.33e+06 & -3.21e-05 & 0.0665 & -7.07 & 1220.1 & 1185.7 \\
QME & 4625.21 & -191.29 & 0.0594 & 2.34e+06 & 1.65e-05 & 22.6523 & -10.67 & 1227.2 & 1192.7 \\
QMICS & 1183.05 & 2239.47 & 0.0682 & 2.34e+06 & 1.91e-05 & 19.7482 & -10.84 & 1227.6 & 1193.1 \\
QMIEU & -2045.21 & -7211.07 & 0.0499 & 2.34e+06 & -2.44e-06 & 26.5579 & -10.53 & 1227.0 & 1192.5 \\
\midrule

\multicolumn{10}{l}{\textbf{SK-I (Raw)}} \\
\midrule
LS\_Standard & 2843.81 & 316.41 & 9.4057 & 2.34e+06 & -1.39e-05 & 0.1059 & 0.11 & 416.0 & 386.6 \\
GLS & -7663.72 & 25896.06 & 9.4047 & 2.35e+06 & -1.56e-04 & 0.1059 & 0.11 & 415.9 & 386.5 \\
BLS & 13353.62 & -9385.32 & 9.4095 & 2.34e+06 & -3.26e-06 & 0.1066 & 0.11 & 415.9 & 386.5 \\
LKSL & 22033.21 & 17756.76 & 9.4100 & 2.34e+06 & -7.52e-07 & 0.0784 & 0.11 & 415.9 & 386.5 \\
PDM1 & -11159.56 & -5887.11 & 9.4006 & 2.35e+06 & -5.24e-05 & 0.1065 & 0.11 & 416.0 & 386.6 \\
MHAOV & 367.17 & 17004.88 & 9.4068 & 2.35e+06 & 1.87e-05 & 0.1061 & 0.11 & 416.0 & 386.6 \\
QME & 386.36 & -3159.93 & 0.3697 & 2.35e+06 & 1.06e-04 & 2.9036 & -5.56 & 427.3 & 397.9 \\
QMICS & -4192.51 & 11580.58 & 9.4033 & 2.35e+06 & -4.40e-05 & 0.1059 & 0.11 & 415.9 & 386.6 \\
QMIEU & 13897.24 & -6717.76 & 9.4171 & 2.35e+06 & 8.71e-05 & 0.1059 & 0.11 & 415.9 & 386.5 \\
\midrule
\multicolumn{10}{l}{\textbf{SK-I (Modified)}} \\
\midrule
LS\_Standard & -19996.40 & -35548.39 & 9.4142 & 2.35e+06 & 1.18e-04 & 0.1059 & 0.40 & 408.4 & 379.0 \\
GLS & -25277.04 & -19324.45 & 9.4087 & 2.35e+06 & 3.77e-05 & 0.1059 & 0.40 & 408.4 & 379.0 \\
BLS & 23835.62 & -15879.29 & 9.4145 & 2.35e+06 & -1.71e-04 & 0.1066 & 0.40 & 408.4 & 379.0 \\
LKSL & 38885.67 & -17647.14 & 9.4153 & 2.35e+06 & -4.45e-05 & 0.0784 & 0.40 & 408.4 & 379.0 \\
PDM1 & -8482.46 & -7368.49 & 9.4163 & 2.34e+06 & 8.93e-05 & 0.1065 & 0.40 & 408.4 & 379.0 \\
MHAOV & -8680.70 & -9663.04 & 9.4077 & 2.35e+06 & 2.00e-05 & 0.1061 & 0.40 & 408.4 & 379.0 \\
QME & -2614.78 & 6294.42 & 0.1299 & 2.34e+06 & 1.08e-05 & 7.5151 & -8.98 & 427.2 & 397.8 \\
QMICS & -15473.89 & -15005.14 & 9.4252 & 2.34e+06 & -7.20e-06 & 0.1058 & 0.40 & 408.4 & 379.0 \\
QMIEU & -27551.47 & -5524.11 & 9.4128 & 2.35e+06 & 1.10e-04 & 0.1058 & 0.40 & 408.4 & 379.0 \\
\midrule

\multicolumn{10}{l}{\textbf{SK-II (Raw)}} \\
\midrule
LS\_Standard & 10907.94 & 25731.16 & 0.8858 & 2.42e+06 & 4.10e-05 & 1.1371 & -1.96 & 191.0 & 165.2 \\
GLS & -7084.36 & 12797.06 & 9.2050 & 2.40e+06 & -1.89e-05 & 0.0565 & -4.63 & 196.4 & 170.6 \\
BLS & 12160.98 & 13686.68 & 0.9026 & 2.42e+06 & -1.65e-04 & 0.7586 & -1.96 & 191.0 & 165.1 \\
LKSL & 12928.23 & -4092.77 & 0.8881 & 2.41e+06 & 3.39e-05 & 1.0782 & -1.96 & 191.0 & 165.1 \\
PDM1 & -6799.31 & -28270.51 & 9.1881 & 2.39e+06 & -1.38e-04 & 0.0664 & -4.62 & 196.3 & 170.5 \\
MHAOV & -7279.82 & 10413.89 & 9.2145 & 2.39e+06 & 7.96e-06 & 0.0565 & -4.62 & 196.3 & 170.5 \\
QME & 48620.28 & -25765.43 & 0.8809 & 2.42e+06 & -2.23e-05 & 1.1598 & -1.96 & 191.0 & 165.1 \\
QMICS & -2495.59 & 2047.89 & 9.2131 & 2.40e+06 & 5.15e-05 & 0.0564 & -4.62 & 196.4 & 170.6 \\
QMIEU & -19493.21 & 7924.27 & 9.2253 & 2.39e+06 & 6.64e-05 & 0.0564 & -4.62 & 196.3 & 170.5 \\
\midrule
\multicolumn{10}{l}{\textbf{SK-II (Modified)}} \\
\midrule
LS\_Standard & 9616.58 & -1140.07 & 9.2071 & 2.40e+06 & 2.74e-05 & 0.0666 & -4.53 & 191.6 & 165.8 \\
GLS & -16100.25 & -17088.47 & 9.2377 & 2.39e+06 & 2.32e-05 & 0.0565 & -4.53 & 191.6 & 165.8 \\
BLS & 1023.07 & -2685.44 & 0.9051 & 2.40e+06 & 6.41e-05 & 0.7586 & -4.29 & 191.2 & 165.3 \\
LKSL & 14086.47 & 1721.68 & 4.5166 & 2.39e+06 & 4.84e-05 & 0.2266 & -4.46 & 191.5 & 165.7 \\
PDM1 & 3893.36 & 12479.22 & 9.1959 & 2.40e+06 & -2.01e-05 & 0.0664 & -4.53 & 191.7 & 165.9 \\
MHAOV & -11505.23 & 502.24 & 9.2128 & 2.40e+06 & -8.29e-06 & 0.0565 & -4.53 & 191.6 & 165.8 \\
QME & -10161.22 & -4152.68 & 0.8595 & 2.41e+06 & 6.17e-05 & 1.1796 & -4.29 & 191.1 & 165.3 \\
QMICS & -6719.23 & 9645.97 & 9.2298 & 2.39e+06 & -3.83e-05 & 0.0564 & -4.53 & 191.7 & 165.8 \\
QMIEU & -8082.64 & -14549.71 & 9.1972 & 2.39e+06 & 3.10e-05 & 0.0564 & -4.53 & 191.7 & 165.9 \\
\midrule

\multicolumn{10}{l}{\textbf{SK-III (Raw)}} \\
\midrule
LS\_Standard & -13158.60 & -39306.77 & 1.7330 & 2.39e+06 & 1.00e-04 & 0.5271 & -2.65 & 186.7 & 161.9 \\
GLS & 11108.77 & -2207.40 & 1.7310 & 2.40e+06 & 1.25e-04 & 0.5562 & -2.65 & 186.7 & 162.0 \\
BLS & 21877.12 & 3478.85 & 1.7469 & 2.39e+06 & -2.05e-05 & 0.5094 & -2.65 & 186.7 & 162.0 \\
LKSL & -16121.71 & 1628.27 & 4.4653 & 2.39e+06 & 2.48e-05 & 0.2805 & -5.91 & 193.2 & 168.5 \\
PDM1 & -32258.82 & -9022.48 & 9.4278 & 2.38e+06 & -1.36e-04 & 0.0931 & -3.08 & 187.5 & 162.8 \\
MHAOV & -20728.85 & -1120.66 & 1.7461 & 2.40e+06 & -5.07e-05 & 0.5625 & -2.65 & 186.7 & 161.9 \\
QME & 13191.42 & 17629.75 & 1.7269 & 2.40e+06 & -8.26e-05 & 0.5993 & -2.65 & 186.7 & 161.9 \\
QMICS & 26606.72 & -31780.91 & 1.7237 & 2.40e+06 & -1.25e-04 & 0.6029 & -2.65 & 186.7 & 161.9 \\
QMIEU & 13991.31 & -14351.55 & 1.7045 & 2.39e+06 & 6.23e-05 & 0.6029 & -2.65 & 186.7 & 161.9 \\
\midrule
\multicolumn{10}{l}{\textbf{SK-III (Modified)}} \\
\midrule
LS\_Standard & -8941.81 & 24903.04 & 1.0748 & 2.39e+06 & -1.43e-05 & 1.1063 & -4.71 & 195.8 & 171.1 \\
GLS & 31832.89 & -13404.53 & 1.7236 & 2.39e+06 & -2.62e-05 & 0.5562 & -3.52 & 193.5 & 168.7 \\
BLS & 1518.31 & -14406.25 & 1.6970 & 2.39e+06 & 1.74e-04 & 0.5094 & -3.52 & 193.5 & 168.7 \\
LKSL & 1361.87 & -5803.31 & 4.5041 & 2.38e+06 & -1.96e-06 & 0.2805 & -6.50 & 199.5 & 174.7 \\
PDM1 & -20885.11 & -1713.84 & 9.4755 & 2.38e+06 & 2.41e-05 & 0.0584 & -3.28 & 193.0 & 168.2 \\
MHAOV & -24775.94 & -3049.75 & 1.7024 & 2.39e+06 & 4.29e-05 & 0.5563 & -3.52 & 193.5 & 168.8 \\
QME & -11964.58 & -17607.21 & 1.6902 & 2.39e+06 & 2.41e-05 & 0.5993 & -3.52 & 193.5 & 168.8 \\
QMICS & -48857.55 & 37676.95 & 1.6838 & 2.39e+06 & -6.04e-05 & 0.6139 & -3.52 & 193.5 & 168.7 \\
QMIEU & 584.69 & 36878.67 & 1.6945 & 2.39e+06 & -3.25e-05 & 0.6102 & -3.52 & 193.5 & 168.8 \\
\midrule

\multicolumn{10}{l}{\textbf{SK-IV (Raw)}} \\
\midrule
LS\_Standard & -6087.22 & -7873.14 & 1.1936 & 2.33e+06 & 1.11e-04 & 0.9530 & 0.43 & 860.3 & 827.7 \\
GLS & 7757.74 & -9121.21 & 1.0506 & 2.33e+06 & -1.39e-04 & 0.9530 & 0.43 & 860.3 & 827.8 \\
BLS & -7337.49 & 15155.86 & 0.3021 & 2.33e+06 & 1.95e-05 & 2.1162 & -4.75 & 870.7 & 838.1 \\
LKSL & 3714.06 & 4498.58 & 9.4438 & 2.33e+06 & -6.62e-05 & 0.0665 & -5.87 & 872.8 & 840.3 \\
PDM1 & -2916.61 & 1153.06 & 9.4520 & 2.33e+06 & -2.06e-05 & 0.0657 & -5.88 & 872.9 & 840.3 \\
MHAOV & -2094.67 & -11109.99 & 1.0424 & 2.33e+06 & 2.44e-05 & 0.9602 & 0.43 & 860.3 & 827.7 \\
QME & -1411.93 & 474.28 & 9.4545 & 2.33e+06 & 7.77e-05 & 0.0657 & -5.87 & 872.8 & 840.3 \\
QMICS & -1847.81 & -8030.20 & 9.4515 & 2.33e+06 & -2.51e-05 & 0.0657 & -5.87 & 872.9 & 840.4 \\
QMIEU & -3500.50 & 2325.11 & 9.4528 & 2.33e+06 & 7.09e-05 & 0.0657 & -5.88 & 872.9 & 840.4 \\
\midrule
\multicolumn{10}{l}{\textbf{SK-IV (Modified)}} \\
\midrule
LS\_Standard & 10239.02 & 1919.85 & 1.6980 & 2.33e+06 & 1.23e-06 & 0.5978 & -2.82 & 857.3 & 824.8 \\
GLS & -965.23 & -1813.59 & 9.4505 & 2.32e+06 & 6.48e-07 & 0.0658 & -5.85 & 863.3 & 830.7 \\
BLS & -4169.74 & -5743.44 & 0.3062 & 2.32e+06 & 6.28e-05 & 2.1162 & -4.83 & 861.2 & 828.6 \\
LKSL & -2911.00 & 1487.40 & 9.4580 & 2.32e+06 & 9.50e-05 & 0.0665 & -5.85 & 863.3 & 830.7 \\
PDM1 & -563.89 & -1018.61 & 9.4506 & 2.32e+06 & 4.59e-05 & 0.0658 & -5.85 & 863.3 & 830.8 \\
MHAOV & -807.37 & -1761.03 & 9.4428 & 2.32e+06 & -4.51e-05 & 0.0658 & -5.86 & 863.3 & 830.7 \\
QME & -191.67 & 980.64 & 1.7122 & 2.32e+06 & -2.16e-04 & 0.5333 & -2.82 & 857.3 & 824.8 \\
QMICS & 2828.47 & 564.76 & 9.4583 & 2.32e+06 & 1.10e-04 & 0.0657 & -5.86 & 863.3 & 830.8 \\
QMIEU & 3197.41 & -1479.82 & 9.4579 & 2.32e+06 & 2.26e-05 & 0.0657 & -5.85 & 863.2 & 830.7 \\
\end{longtable}
\endgroup

\begin{tablenotes}
\small
\item \textbf{Note:} This table combines best-fit parameters (left) and summary statistics (right). All fits use the bin-averaged sinusoidal model with asymmetric Gaussian likelihood. $\ln \mathrm{B}$ is the logarithmic Bayes factor via BIC approximation (Eq.~3.27); evidence strength evaluated with Jeffreys' scale. Flux parameters in raw units; $\nu$ in yr$^{-1}$; slope in day$^{-1}$. 
\end{tablenotes}

\bibliographystyle{JHEP}
\bibliography{references.bib}

\end{document}